\documentclass[12pt]{article}
 \usepackage{ifpdf}
 \usepackage{epsfig}
  \usepackage[latin1]{inputenc}
 \usepackage{amsfonts}
 \usepackage{amsthm}
 \usepackage{amssymb, latexsym, amsmath}
 \usepackage{amsbsy}
\usepackage{amstext}
\usepackage{amsopn}
\usepackage{amsgen}
\usepackage[dvips]{color}
 \usepackage{graphicx}
  \usepackage{color}
  \usepackage{subfigure}
 \newcommand{\be}{\begin{equation}}
 \newcommand{\ee}{\end{equation}}
 \newcommand{\ba}{\begin{eqnarray}}
 \newcommand{\ea}{\end{eqnarray}}

 \newcommand{\erbold}{\mbox{{\boldmath $\vec r$}}}

\begin{document}
  \title{
         ${{~~~~~~~~~~~~~~~~~~~~~~~~~~}^{^{^{
         Published~in~the~
                 Astrophysical~Journal
        \,,~Vol.~764\,,~id.~27\,~~(2013)
                  }}}}$\\
 {\Large{\textbf{{{No pseudosynchronous rotation for terrestrial planets and moons}
  ~\\
  ~\\}
            }}}}
 \author{
  {\Large{Valeri V. Makarov}}\\
  {\small{US Naval Observatory, Washington DC 20392}}\\
 {\small{e-mail: ~vvm @ usno.navy.mil~}}\\
     ~\\
     {\Large{and}}\\
     ~\\
   {\Large{Michael Efroimsky}}\\
 {\small{US Naval Observatory, Washington DC 20392}}\\
 {\small{e-mail: ~michael.efroimsky @ usno.navy.mil~}}\\
  }
     \date{}

 \maketitle

 \begin{abstract}
 We reexamine the popular belief that a telluric planet or satellite on an eccentric orbit can, outside a spin-orbit resonance, be captured in a quasi-static tidal
 equilibrium called pseudosynchronous rotation. The existence of such configurations was deduced from oversimplified tidal models assuming either a constant tidal
 torque or a torque linear in the tidal frequency. A more accurate treatment requires that the torque be decomposed into the Darwin-Kaula series over the tidal modes,
 and that this decomposition be combined with a realistic choice of rheological properties of the mantle, which we choose to be a combination of the Andrade model at
 ordinary frequencies and the Maxwell model at low frequencies. This development demonstrates that there exist no stable equilibrium states for solid planets and moons,
 other than spin-orbit resonances.
 \end{abstract}


\section{Motivation}

 The ongoing quest for extraterrestrial life has placed exoplanets and their properties into the forefront of scientific investigation.
 The trend has provided additional momentum to the development of a broad variety of techniques and approaches employed in the planetary
 sciences. For example, the recent revival of interest in mechanics of bodily tides is partly due to the importance of the planetary
 spin for the prospects of finding habitable worlds near other stars.

 Well-known examples of dynamical equilibria achieved via tidal coupling include our Moon, which is in a 1:1 spin-orbit
 resonance to the Earth, and Mercury which makes exactly three sidereal rotations over every two orbital revolutions around the Sun.
 Similar behaviour is expected of the growing number of known super-earths -- especially if their composition happens to be similar to
 that of the terrestrial planets of the solar system, i.e., if they have massive solid or partially molten mantles of rocky minerals.

 Unfortunately, some of the published far-reaching conclusions about specific exoplanets are based on incomplete or $\,${\it{ad hoc}}$\,$ models which
 should never be used for solid materials, including those with partial melt.
 Both these models, introduced by Goldreich (1966) mainly for the ease of analytical treatment, predict quasi-static
 pseudosynchronous rotation states, with the planet being trapped in a slowly changing equilibrium state at a faster-than-synchronous rotation rate and a vanishing orbit-averaged tidal torque. Except in specific (very narrow) frequency bands, these
 models are incompatible with the rheological properties of realistic mantles and crusts. Analysis based on actual rheologies
 demonstrates the impossibility of pseudosynchronous rotation for homogeneous terrestrial objects. Whether this prohibition extends
 to planets and moons with internal or surface oceans remains an open issue and needs further research.

 \section{The constant angular lag model}

 A consistent linear theory of bodily tides is based on Fourier decomposition of the tide, with subsequent inclusion of the response at each separate mode.
 The ensuing level of complexity has tempted many to circumvent it by developing simpler approaches. Serving as good illustrations and reflecting some qualitative aspects of the tidal interaction, such models are not necessarily applicable for quantitative purposes (Efroimsky \& Lainey 2007) and should certainly be eschewed when fine features of
 near-resonant dynamics are explored.

 \subsection{The essence of the method}

 One, often-used, toy model prescribes:\\

 (a) to set both the Love number $\,k_2\,$ and geometric lag $\,\epsilon_g\,$
 frequency-independent;\\
 ~\\

 (b) to insert their values into the popular short formula
 \begin{subequations}
 \ba
 {\cal{T}}_z
 ~=~\frac{3}{2}~G\,M_{1}^{\,2}~\frac{R^5}{r^6}~k_2\,\sin2\epsilon_g
 \label{1a}
 \ea
 $~~~~~~~~~~$for the polar component of the torque wherewith a tide-raising perturber of mass $\,M_{1}\,$\\
 $~~~~~~~~~~~$acts on a tidally perturbed body of radius $\,R\,$ located at a distance $\,r\,$,  the
 obliquity $\,i\,$\\
 $~~~~~~~~~~~$assumed small; and\\
 ~\\


 (c)~ to combine formula (\ref{1a}) with the assumption that the angle $\,\epsilon_g\,$ stays constant while\\
 $~~~~~~~~~~~~$the tide-raising perturber stays on one side of the bulge (in the sense of the directions\\
 $~~~~~~~~~~~~$as seen from the perturbed body's centre).\\
 ~\\

 \noindent
 For a nonzero eccentricity $\,e\,$, and in a sufficient proximity of the 1:1 resonance, the relative orientation of the perturber and the
 bulge changes twice over an orbital period. Hence, within this model, the angle $\,\epsilon_g\,$ is set, by hand, to
 change its sign abruptly twice in a cycle, while keeping its magnitude $\,|\,\epsilon_g\,|\,$ fixed. Therefore the model can be written
 down as
 \ba
 {\cal{T}}_z
 ~=~\frac{3}{2}~G\,M_{1}^{\,2}~\frac{R^5}{r^6}~k_2\,\sin2|\epsilon_g|~\,\mbox{Sgn}\,(\dot{\nu}\,-\,\dot{\theta})\,~,
 \label{1b}
 \ea
 \label{1}
 \end{subequations}
 $\theta$ and $\dot{\theta}$ being the perturbed body's sidereal angle and spin rate, and $\nu$ being the true anomaly.

 Historically, the method dates back to the paper by MacDonald (1964) and therefore is often referred to as the {\it{MacDonald torque}}
 (e.g., Touma \& Wisdom 1994, Section 2.7.1).

 The approach is also called the $\,${\it{constant angular lag model}}$\,$ or the $\,${\it{constant tidal torque model}}, both names being
 somewhat misleading. Indeed, in the vicinity of the 1:1 resonance the sign of the lag (and the torque) is set positive
 or negative, when the bulge falls behind or advances relative to the direction towards the perturber. So both quantities change their
 sign twice over a period -- a circumstance that makes the term {\it{constant}} inappropriate. Furthermore, the torque depends upon the
 distance and is always evolving in time unless the orbit is circular.

 The abrupt switch of the sign of the torque (\ref{1}), with its magnitude staying unchanged, is quite a contrived
 assertion$\,$\footnote{~Stated alternatively, if we represent $\,(a/r)^6\,$ as a series of Fourier harmonics $\cos(j{\cal{M}})$,
 $j=0,1,\ldots$, where $a$ is the semimajor axis and ${\cal{M}}$ is the mean anomaly, we shall have to accept that the $\cos({\cal{M}})$
 tidal mode generates a positive (accelerating) torque for ${\cal{M}}\in [-\pi/2,+\pi/2]$, abruptly switching to a negative value for
 ${\cal{M}}\in [\pi/2,3\pi/2]\,$.} which by itself indicates that the model is unphysical. A deeper, mathematical, objection will be
 brought up in Section \ref{damn}.

 Saying goodbye to the constant-angular-lag
 model will not be easy, because it has been a textbook standard for nearly half a century. Given the
 attractive simplicity of the model, one will always be tempted to enquire if perhaps it would still be producing at least qualitative results of some value. To see
 that it would not, we shall have to scrutinise the principal outcomes of the model.

 The perturber's orbit is set to lie in
 the equatorial plane of the perturbed body; in other words, the obliquity is set zero. Two special situations of interest emerge here. One is the case of exact synchronism, the other being the case of a vanishing average tidal torque. Both settings were explored by Goldreich (1966) whose results are explained in detail
 by Murray \& Dermott (1999).

 \subsection{The synchronous spin case}\label{sub}

 Suppose the tidally-perturbed body on an elliptic ($\,e\,\neq\,0\,$) orbit is caught into the 1:1 spin-orbit resonance: $\,n=
 \dot{\theta}\,$. Then, as explained in Section 5.2 of Murray \& Dermott (1999), the angular motion rate $\,\dot{\theta}\,$ exceeds $\,n\,$
 over exactly one half of the orbital time period, and falls short of $\,n\,$ during the other half of the period.
 Correspondingly, the tidal torque $\,{\cal{T}}_z
 \,$ is positive (accelerating) through the former half of the period, and is negative
 (decelerating) through the latter half. The instantaneous tidal torque is proportional to a negative power of the instantaneous
 distance $\,r\,$ between the bodies. As depicted in Figure 5.3 in {\it{Ibid}}, when the disturbed body's angular motion is faster than
 the mean motion, the bodies are closer, so the positive (spinning up) tidal torque is larger in absolute value than the negative torque
 for the other half of the period.
 %
 %
 Thus the resultant orbit-averaged torque $\,\langle\,{\cal{T}}_z
 \,\rangle\,$ is positive, and the net effect is to accelerate the
 tidally perturbed body's spin. (Recall that the undisturbed body is assumed to be spherical or oblate, so the tidal torque is the only
 one coming into play.)

 \subsection{The case of vanishing tidal torque}

 The second
 important application
 of the constant angular lag model is the situation where the
 orbit-average tidal torque vanishes: $\,\langle\,{\cal{T}}_z
 \,\rangle\,=0\,$. Vanishing of the average tidal torque entails a dynamical equilibrium: the tidally disturbed body
 keeps rotating at a steady spin rate. A calculation of this rate, borrowed from Goldreich (1966), is presented in Murray \& Dermott
 (1999) and is often cited in the literature. According to that development, the equilibrium is achieved, for a zero obliquity, at the
 spin rate of
 \begin{equation}
 \stackrel{\mbox{\bf{\LARGE{$\,\cdot$}}}}{\theta}_{\textstyle{_{\rm eq}}}\,=\,n\,\left(\,1\,+\,\frac{19}{2}~e^2\,\right)~\,,
 \label{6}
 \end{equation}
 $e\,$ being the eccentricity. At first glance, the result looks unassailable. Indeed, for $\,\dot{\theta}=n\,$, the bulge is lagging
 behind the central line over one half of the time period (around the periastron), so the torque accelerates the rotation. Over the other
 half of the period, the torque decelerates, being weaker due to a larger distance. So the state $\,\dot{\theta}=n\,$ looks
 unstable, as the overall average torque seems to be accelerating.

 It is however well known that the Moon is not staying in this pseudosynchronous regime (which would be 3\% faster than the
 synchronous rotation wherein the Moon is presently locked). Murray \& Dermott (1999) point at the lunar quadrupole moment as the
 reason why the Moon is not pseudosynchronous. A deeper reason though lies in the constant geometric lag model being genuinely flawed,
 and in the entire calculation leading to (\ref{6}) being invalid.

 \subsection{A major objection against the constant-angular-lag model}\label{damn}

 As well known, the generic expression for the tidal amendment to the perturbed body's potential is furnished by a Fourier series
 developed by Kaula (1964). We term it {\it{the Darwin-Kaula expansion}}, as a partial sum of that series was written down by Darwin
 (1879). Accordingly, the generic expression of the tidal torque also must look as an infinite series. The series remains infinite even
 if we include into it only the degree-2 terms, i.e., those proportional to the quadrupole Love number $\,k_2\,$. The very fact that the
 expansion for the torque can be wrapped into a short and neat form (\ref{1}) is an indicator of some extra, very special assumption
 being involved.

 As was pointed out in Williams \& Efroimsky (2012), such an assumption indeed is present in the constant-angular-lag
 model, though this assumption is never stipulated explicitly. The situation is elucidated in all detail in the paper by Efroimsky \&
 Makarov (2013) to which we refer the reader. Here we shall provide only a brief summary.

 As explained in Williams \& Efroimsky (2012), the afore-presented concise expression (\ref{1}) for the torque is equivalent to the full Darwin-Kaula expansion for the
 potential, only if the following assumptions are made:

 \begin{itemize}

 \item{} In all terms of the Darwin-Kaula series for the tidal amendment to the potential of the perturbed body, i.e, for all
         Fourier tidal modes $\,\omega_{lmpq}\,$, the time lags are endowed with the same, frequency-independent value $\,\Delta t\,$.

 \item{} The obliquity is set small.

 \item{} Only the terms with $\,l=m=2\,$ are retained, and the Love number $\,k_2\,$ entering these terms is assumed
 frequency-independent. Here the degree $\,l\,$ and the order $\,m\,$ are the first two integers of the four-number set $\,lmpq\,$
 used to number the Fourier modes showing up in the Darwin-Kaula expansion.\\

 \end{itemize}

 \noindent
 The so-processed Darwin-Kaula series for the tidal potential becomes equivalent to a concise expression (equation 16b in {\it{Ibid}}.)
 wherefrom our expression (\ref{1}) for the torque ensues.

 Alternatively, the above three assumptions could be applied directly to the Darwin-Kaula expansion for the tidal torque. Once again, the outcome would be the above expression (\ref{1}) for the torque. This is demonstrated in Efroimsky \& Makarov (2013, equation 34).

 Under the three assumptions, the instantaneous geometric lag angle turns out to be \footnote{~The geometric angular lag $\epsilon_g$
 is not to be confused with the instantaneous {\it{phase}} lag (or longitudinal lag)
 \ba
 \epsilon_{ph}\,\equiv~2\,(\dot{\nu}-\dot{\theta})\,\Delta t~=~2~\epsilon_g
 \nonumber
 \ea
 sometimes used in the literature (Efroimsky \& Williams 2009, Williams \& Efroimsky 2012).}
 \ba
 \epsilon_g\,\equiv~(\dot{\nu}-\dot{\theta})\,\Delta t~\,,
 \label{2}
 \ea
 $\,\Delta t\,$ being the frequency-independent time lag, $\,\nu\,$ being the true anomaly of the perturber, and $\,\theta\,$ being the sidereal angle of the
 tidally perturbed body. From this expression, it is straightforward that the validity of formula (\ref{1}) is incompatible with the geometric lag being
 constant. Indeed, the road to (\ref{1}) is paved with the aforementioned three assumptions, one of which being that of a constant $\,\Delta t\,$. As can be
 observed from (\ref{2}), the latter is incompatible with the geometric lag being constant, unless the eccentricity is nil. \footnote{~As a last resort, one
 can suggest (a) to tune the time dependence of $\,\Delta t\,$ so that the lag angle $\,\epsilon_g\,$ in (\ref{2}) stays constant in time, and (b) to assume
 that the time lags at all Fourier tidal modes are equal to the so specially evolving $\,\Delta t\,$. This would imply that the lagging properties of the
 material {\it{at all frequencies}} are being tuned in a fine manner, continuously and simultaneously, so that $\,\Delta t\,$ adjusts its evolution rate, to
 stay inverse to $\,\dot{\nu} -\dot{\theta}\,$ at any instant of time. The rate of change of $\,\dot{\nu} -\dot{\theta}\,$ being defined by the orbital
 parameters, existence of such a rheology in nature looks impossible.}

 On all these grounds, the constant-geometric-lag (constant-torque) model should be discarded as such.

 \section{Pseudosynchronism in the constant time lag model}

 As distinct from the constant geometric lag approach, the constant {\it{time}} lag model sets the time delay $\,\Delta t\,$ independent
 of the tidal mode frequency. Pioneered by Darwin (1879), this assumption was a part of numerous works, e.g., Hut (1981), Eggleton et al.
 (1998). The assumption was also the base for one of the two models considered by Goldreich \& Peale (1966, equation 23), the other model
 addressed in that paper being the afore-discarded constant geometric lag method.

 The constant time lag model is unique, in that it makes the full Darwin-Kaula expansion for the tidal potential (or torque) equivalent
 to a much shorter and simpler expression. In regard to the tidal potential, this is the equivalence of the full series (16) and a simpler
 expression (15) in our preceding paper Efroimsky \& Makarov (2013). In application to the torque, this is the equivalence of the
 appropriate full series to the simpler expression (34) in {\it{Ibid}}.

 If we agree to limit our approximation to the lowest degree and order, $\,l=m=2\,$, the aforementioned simpler expressions read as
 \ba
 U(\erbold)=~-~\frac{3}{4}~G\,M_{1}\,k_{2}~\frac{\,R^{\textstyle{^5}}\,}{\,r^6\,}~\cos (\,2\,(\dot{\nu}-\dot{\theta})\,\Delta t\,)\,=~
 \frac{3}{4}~G\,M_{1}\,k_{2}~\frac{\,R^{\textstyle{^5}}\,}{\,r^6\,}~\cos (\,2\,|\,\dot{\nu}-\dot{\theta}\,|\,\Delta t\,)
 \,~,\,~\,
 \label{MacDonald_1_b}
 \ea
 for the potential, and as
 \ba
 {\cal{T}}_z
 \,=\,\frac{\textstyle 3}{\textstyle 2}~{G\;M_{1}^{\,2}}\,k_{2}\;\frac{R^5}{r^{
 \textstyle{^{6}}}
 }\;\sin (\,2\,(\dot{\nu}-\dot{\theta})\,\Delta t\,)~=~
 \frac{\textstyle 3}{\textstyle 2}~{G\;M_{1}^{\,2}}\,k_{2}\;\frac{R^5}{r^{
 \textstyle{^{6}}}
 }\;\sin (\,2\,|\,\dot{\nu}-\dot{\theta}\,|\,\Delta t\,)~\,\mbox{Sgn}\,(\dot{\nu}-\dot{\theta})
 \,~,\quad
 \label{L22}
 \ea
 for the polar torque.
 A detailed derivation of (\ref{MacDonald_1_b}) can be found in Williams \& Efroimsky
 (2012), while derivation of (\ref{L22}) is offered in Efroimsky \& Makarov (2013).

 To average the torque, it is instrumental to insert into (\ref{L22}) the distance $\,r\,$ expressed through the semimajor axis $\,a\,$,
 eccentricity $\,e\,$, and true anomaly $\,\nu\,$, and to integrate over the orbital cycle. The procedure gets simplified greatly for
 small lags, when one can substitute the sine with its argument. Then the calculation (presented in detail in the Appendix to Williams
 \& Efroimsky 2012) renders:
 \begin{equation}
 \langle\,{\cal{T}}_z
 \,\rangle~\propto ~
 \left[ \frac{~1~+~\frac{\textstyle 15}{\textstyle 2}~e^2\,+~\frac{\textstyle 45}{\textstyle 8}~e^4\,+~\frac{\textstyle 5}{\textstyle
 16}~e^6~}{(1~-~e^2)^{\textstyle{^{6}}}}~-~\frac{\,\dot\theta\,}{\,n\,}~\frac{~1~+~3~e^2\,+~\frac{\textstyle 3}{\textstyle 8}~e^4~}{
 (1~-~e^2)^{\textstyle{^{9/2}}}}\right]~\,.
 \label{avertorque}
 \end{equation}
 This expression was obtained by Eggleton et al. (1998), though its equivalent was present in an earlier paper by Hut (1981, equation 11).
 In a somewhat disguised form, this expression can be found in a much earlier work by Goldreich \& Peale (1966, equation 24).

 We find readily that the equilibrium (i.e., vanishing of the average tidal torque) is achieved at
 \begin{equation}
 \dot\theta_{\rm
  equ
 }~=~n~\left[\,1\,+\,6\,e^2\,+\,
 \frac{3}{8}
 ~e^4\,+\,
 \frac{173}{8}
 ~e^6\,+\,O(e^8)\,\right]~\,.
 \label{above}
 \end{equation}

\begin{figure}
 \includegraphics[width=159mm]{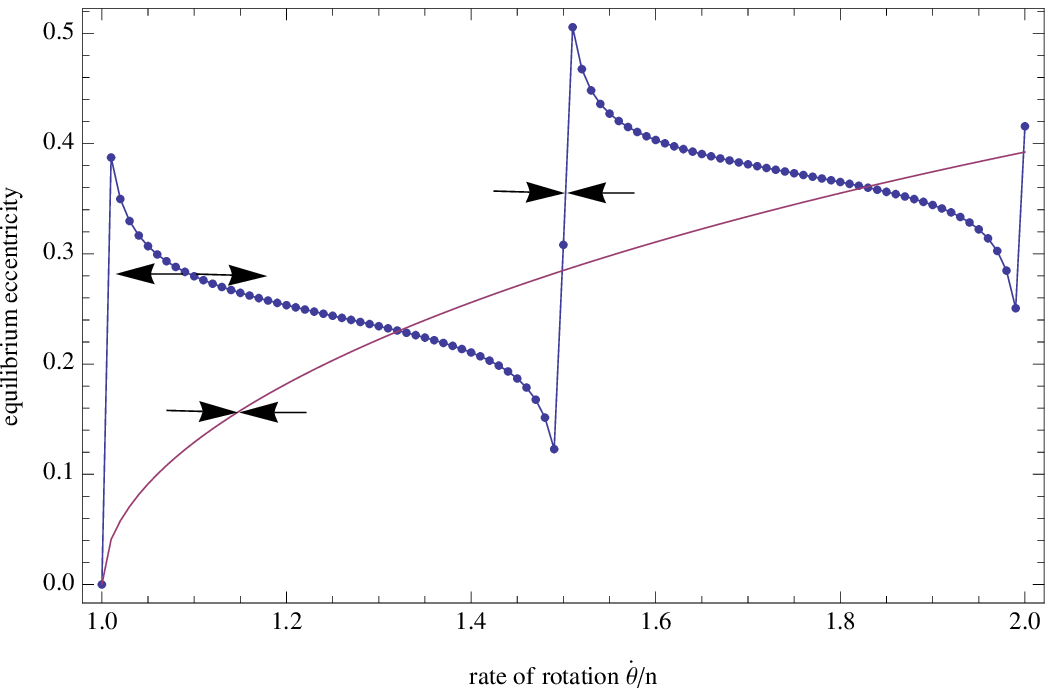}
 \caption{\small{Equilibrium eccentricity (one corresponding to a vanishing average tidal torque) depicted against the dimensionless spin
 rate $\,\dot\theta/n\,$. Calculations were made for a tidally perturbed rotating body with parameters of the Moon, as shown in Table
 \ref{table}.
 $~\quad\quad\quad\quad\quad\quad\quad\quad\quad\quad\quad\quad\quad\quad\quad\quad\quad\quad\quad\quad\quad\quad\quad\quad\quad\quad
 \quad\quad\quad\quad\quad\quad\quad\quad\quad\quad\quad\quad\quad\quad\quad\quad~$
 The monotonically rising curve corresponds to the linear torque (constant time lag) model. The jagged
 dotted line corresponds to a realistic rheology introduced in Section \ref{eq}. Both functions were computed with a step of $\,0.01\,$ in $\,\dot\theta/n\,$.
 $~\quad\quad\quad\quad~~\quad\quad\quad\quad~~\quad\quad\quad\quad~~\quad\quad\quad\quad~~\quad\quad\quad\quad~~\quad\quad\quad\quad~
 ~\quad\quad\quad\quad~~\quad\quad\quad\quad~~\quad\quad\quad\quad~~\quad\quad\quad\quad~~\quad\quad\quad\quad~~\quad\quad\quad\quad~$
 In both cases, the resulting curve divides the plane into two parts corresponding to the two opposite signs of the average
 polar torque $\,\langle\,{\cal{T}}_z
 \,\rangle\,$. While $\,\langle\,{\cal{T}}_z
 \,\rangle\,$ is positive (accelerating) everywhere
 above the curve, it stays negative (decelerating) everywhere below the curve.
 $~\quad\quad\quad\quad~~\quad\quad\quad\quad~~\quad\quad\quad\quad~~\quad\quad\quad\quad~~\quad\quad\quad\quad~~\quad\quad\quad\quad~
 ~\quad\quad\quad\quad~~\quad\quad\quad\quad~~\quad\quad\quad\quad~\quad\quad\quad\quad~~\quad\quad\quad\quad~~\quad\quad\quad\quad~$
 The small
 arrows indicate the action of the tidal torque upon small perturbations of the spin rate away from an equilibrium state. For the
 constant time lag model, the torque is restoring, and the equilibrium is stable. In the case of realistic rheology, though, the
 emerging nonzero torque drives the rotator away from the stable spin.
 }}
 \label{moon.fig}
 \end{figure}

\begin{figure}
 \includegraphics[width=159mm]{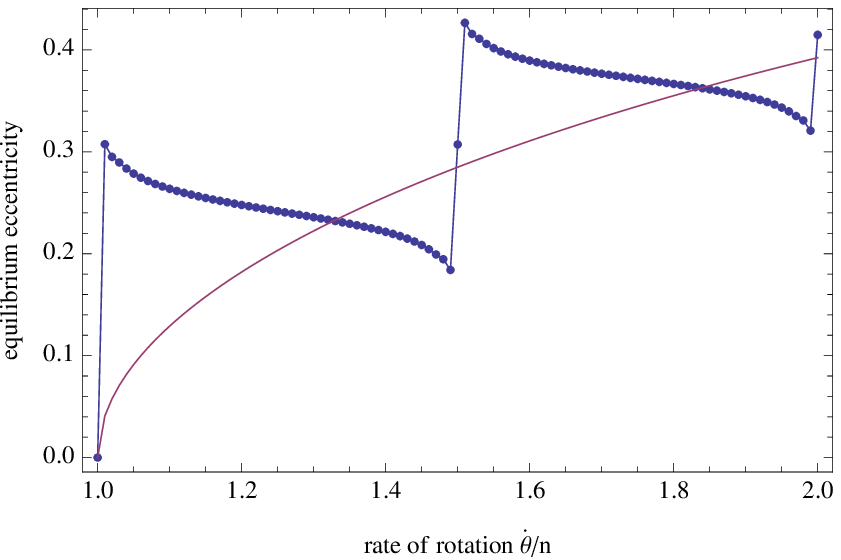}
 \caption{\small{Equilibrium eccentricities of a zero secular tidal torque acting on a tidally perturbed super-Earth, depicted against the dimensionless spin rate
 $\,\dot\theta/n\,$. Parameters of the super-Earth are given in Table \ref{table}, and are consistent with those chosen for GJ581d in Makarov et al. (2012).
 $~\quad\quad\quad\quad\quad\quad\quad\quad\quad\quad\quad\quad\quad\quad\quad\quad\quad\quad\quad\quad\quad\quad\quad\quad\quad\quad
 \quad\quad\quad\quad\quad\quad\quad\quad\quad\quad\quad\quad\quad\quad\quad\quad~$
 The monotonically rising curve represents the prediction of the constant time lag model.
  $~\quad\quad\quad\quad\quad\quad\quad\quad\quad\quad\quad\quad\quad\quad\quad\quad\quad\quad\quad\quad\quad\quad\quad\quad\quad\quad
 \quad\quad\quad\quad\quad\quad\quad\quad\quad\quad\quad\quad\quad\quad\quad\quad~$
  The jigsaw dotted curve illustrates the prediction of the realistic rheological model described in Section \ref{eq}.
  The function was computed for a grid of points at a step of $\,0.01\,$ in $\,
 \dot{\theta}/n\,$.}
 \label{super.fig}}
 \end{figure}

 Note that the pseudosynchronous rate of rotation depends only on the mean motion and eccentricity. This enables us to solve equation
 (\ref{above}) with respect to $\,e\,$, for a fixed dimensionless spin rate $\,\dot\theta/n\,$. The outcome will be the {\it{equilibrium
 eccentricity}}  $\,e_{equ}\,$, i.e., the eccentricity that ensures the vanishing of the average torque at a certain value of $\,\dot
 \theta/n\,$. In Figures \ref{moon.fig} and \ref{super.fig}, the equilibrium eccentricity $\,e_{equ}\,$ is depicted as a function of $\,\dot\theta/n\,$.
 For the model leading to expression (\ref{avertorque}) for the torque, this is a monotonically rising curve.

 The curve divides the plane into two parts corresponding to the two opposite signs of the average
 polar torque $\,\langle\,{\cal{T}}_z
 \,\rangle\,$. While $\,\langle\,{\cal{T}}_z
 \,\rangle\,$ is positive (accelerating) everywhere above the curve, is stays negative (decelerating) everywhere below the curve. Indeed, if we fix the eccentricity and make
 $\,\dot\theta/n\,$ very large, this will guarantee us that we get into the lower right part of the picture, i.e., below the rising curve.
 In this situation, i.e., for a fixed eccentricity and a sufficiently swift spin, the second term of the torque (\ref{avertorque}) must be
 leading, wherefore the torque must be negative, i.e., despinning. Similarly, by fixing the eccentricity and making the spin rate
 very small, we ensure getting into the upper left part of the picture, and also ensure that the first term in (\ref{avertorque}) is
 leading, so the torque is positive, i.e., accelerating the spin. Since the smoothly rising curve$\,$\footnote{~Here and hereafter, the
 symbol $\,e_{equ}(\dot\theta/n)\,$ implies $\,e_{equ}\,$ as a function of the ratio $\,\dot\theta/n\,$. This is not a product of
 $\,e_{equ}\,$ and $\,\dot\theta/n\,$.} $~e_{equ}(\dot\theta/n)\,$
 corresponds to a zero $\,\langle\,{\cal{T}}_z
 \,\rangle\,$, it is impossible to change the sign of $\,\langle\,{
 \cal{T}}_z
 \,\rangle\,$ without crossing the curve.

 Within the constant time lag model, the function $\,e_{equ}(\dot\theta/n)\,$ being a smoothly rising curve explains the emergence of
 pseudosynchronism. To see this, consider a point on this curve, corresponding to a certain pseudosynchronous state. A small perturbation
 in $\,\dot\theta/n\,$ makes the tidally perturbed body rotate either faster or slower than the pseudosyncronous rate, and a nonzero
 tidal torque emerges. Illustrated by the two counter directed short arrows on the plot in Figure \ref{moon.fig}, the tidal torque is {\it restoring}, in
 that its action is opposite to the sign of perturbation. Thus, the tidal torque will return the perturbed rotator to the equilibrium
 state, i.e., to the initial position on the curve. So the equilibrium is stable. \footnote{~Being stable, the equilibrium is
 {\it{quasi}}-static, in the following
 sense. As the tidal dissipation goes on, the process of despinning continues. The argument $\,\dot\theta/n\,$ slowly decreases, and so
 does the appropriate value of the equilibrium eccentricity. After a perturbation in $e$ from an equilibrium state gives birth to a
 torque, the torque corrects swiftly the spin rate in such a way that the rotator returns to an equilibrium state. However, the
 equilibrium state itself is evolving slowly. In the case of a two-body problem, this evolution is always directed towards the
 configuration with $\,\dot\theta/n\,=\,1\,$ and $\,e\,=\,0\,$. For a viscous body, this was proven by Hut (1981). For a broader class
 of viscoelastic rheologies, the proof was offered by Bambusi and Haus (2012).}

 The constant time lag model ignores the important contribution of rigidity (Segatz et al. 1988) and inelasticity (Karato and Spetzler
 1990) into the tidal response of Earth-like planets. As a result, the model is incapable to account correctly for creep.
 As will be discussed in the following section, the
 above derivation of quasi-stable pseudosynchronism, from the linear torque model, is inapplicable to Earth-like planets with rigid
 mantles. However, in the viscous limit, this model may still be applicable to celestial bodies that do not have appreciable
 rigidity or inelasticity, such as gaseous planets and stars. Observations of binary stars, especially of short-period active stars on
 eccentric orbit, hold the best prospect of proving or disproving the linear torque model for this type of objects (Ferraz-Mello 2012,
 Torres et al. 2010).

 The spin rate of active stars can be inferred from the characteristic periods of photometric variations caused by the passage of large
 spots or groups of spots across the visible disk of the star. The orbital period and the eccentricity are determined from spectroscopic
 radial velocity measurements. We find somewhat conflicting evidence for the existence of pseudosychronism in binary stars. Some stars
 with considerable eccentricities appear to have pseudosynchronous rotation (Hall 1986, Fekel et al. 1998), which is consistent with the
 original prediction by Hut (1981). Other stars clearly rotate faster or slower than the predicted rate (Fekel et al. 1993, Strassmeier
 et al. 2011). Even more puzzling, a significant number of tight binary systems have been found on circularised orbits, albeit spinning
 clearly asynchronously. This fact comes into contradiction with one of the important predictions of the linear torque theory, the one
 that synchronisation (or pseudosyncronisation) of rotation is achieved much sooner than circularisation (P. P. Eggleton 2011, private communication). Thus the
 impression created by the current body of observations is that the constant time lag model is, at least, not universally applicable to
 stars. This should not come as a surprise, because there exist theoretical indications that stars may have magnetic rigidity
 (Williams 2004, 2005, 2006; Ogilvie 2008, Garaud et al. 2010).~\footnote{~Another deviation from the purely viscous model can be caused by the so-called $\,\Lambda-$effect responsible for differential rotation (K{\"{a}}pyl{\"{a}}
 \& Brandenburg 2008, Kichatinov 2005, R\"{u}diger 1989). Turbulent convection generates an extra stress called {\it{Reynolds stress}}. While in a non-rotating
 convection zone this stress can be described as an addition to the viscosity, this can no longer be done when the rotation period becomes comparable to or shorter
 than the convective turnover time. In that situation, a non-viscous input, the so-called $\,\Lambda$-effect, shows up. In its presence, the stress tensor in the
 stellar material will no longer be proportional to the time derivative of the strain tensor, but will contain terms proportional directly to velocity. Thus the purely
 viscous model falls apart, and a frequency-independent time lag is no longer an option.}

 Finally, it should be mentioned that, contrary to a common belief, the purely viscous model does {\it{not}} render a frequency-independent time lag at all
 frequencies. Stated differently, the purely viscous model does {\it{not}} imply that the factors $\,k_l(\omega_{lmpq})\,\sin\epsilon_l(\omega_{lmpq})\,$ are linear
 functions of the tidal mode $\,\omega_{lmpq}\,$ for all values of the mode. It can be demonstrated that this linearity takes place at low frequencies, but gets
 violated at frequencies higher than $\,G\rho^2R^2/\eta\,$, where $\,G$, $\,\rho$, $\,R$, and $\,\eta\,$ are the Newton gravity constant, mean density, radius, and
 the mean viscosity of the perturbed body. We shall address this topic elsewhere.

 \begin{table*}
 \centering
 \begin{minipage}{145mm}
 \caption{Parameters of the tidal model.}
 \begin{tabular}{@{}lrrrr@{}}
 \hline
            &                           &       &              &                      \\
   Name     &  Description~~~~~~~~~~    & Units &         \multicolumn{2}{c}{Values}  \\
            &                           &       &    ~Moon~    &    ~super-Earth      \\
            &                           &       &              &    ~($\,$GJ581d$\,$)~        \\
 \hline
 $\xi$ & \dotfill moment of inertia coefficient &   &
  2/5 & 2/5 \\
 $R$   & \dotfill radius of the perturbed body              & m & $1.737\times 10^6$ & $1.083\times 10^7$ \\
 $M_2$ & \dotfill mass of the perturbed body & kg & $7.3477\times 10^{22}$ &$4.23\times 10^{25}$\\
 $M_1$ & \dotfill mass of the perturbing body & kg & $5.97\times 10^{24}$ &$6.17\times 10^{29}$\\
 $a$ & \dotfill semimajor axis & m & $3.84399\times 10^{8}$ &$3.3\times 10^{10}$ \\
 $n$ & \dotfill mean motion, i.e. $2\pi/P_{\rm orb}$ & yr$^{-1}$ & $84$ &$34.25$ \\
 $e$ & \dotfill orbital eccentricity & & 0.0549 &0.27\\
 $(B-A)/C$ & \dotfill triaxiality & & $2.278\times 10^{-4}$ &$5\times 10^{-5}$ \\
 ${G}$ & \dotfill gravitational constant & m$^3$ kg$^{-1}$ yr$^{-2}$ &
     $66468$ & $66468$ \\
 $\tau_M$ & \dotfill Maxwell time
 & yr &  5 &50\\
 $\mu$ & \dotfill unrelaxed rigidity modulus & Pa
  & $0.8\times10^{11}$  &$0.8\times10^{11}$\\
 $\alpha$ & \dotfill the Andrade parameter & & $0.2$ & $0.2$\\
 \hline
 \label{table}
 \end{tabular}
 \end{minipage}
 \end{table*}

 \section{Equilibrium torques for Earth-like planets and moons}\label{eq}

 To build a consistent theory of bodily tides, one has, first, to decompose the tide into a Fourier series and, second, to attribute to each Fourier
 component its own phase delay and magnitude decrease (the latter being expressed by the Love number appropriate to the said Fourier mode). Development
 of the decomposition technique was started by Darwin (1879) and accomplished in full by Kaula (1964). Attribution of phase delays and Love number values
 to the Fourier modes took much longer time, because of the necessity to explore rheological properties of the mantle at various frequencies. This
 exploration, by both seismological and geodetic methods, has been going on intensively through the past dozens of years. Merger of the Darwin-Kaula
 decomposition technique with the results from solid-Earth rheology is explained in Efroimsky (2012$\,$a). The paper relied on a combined rheological model
 (Andrade at higher frequencies, Maxwell at lower frequencies), because of this model's ability to best match laboratory experiments and both seismic and
 geodetic measurements of dissipation over a range of frequencies in the solid Earth.\footnote{~Motivation for the combined model stems from the mantle being predominantly viscoelastic at frequencies below some threshold, and predominantly inelastic at frequencies above it. As
 explained in Karato \& Spetzler (1990), dissipation above the threshold is dominated by defect unpinning
 (see also Miguel et al. 2002).
 When the frequency descends below the threshold, the effectiveness of this mechanism declines, because the Andrade term in the expression for the complex compliance decreases exponentially. The response of the mantle approaches that of the Maxwell body. So slow processes (like the postglacial rebound) are viscoelastic.

 For Earth's mantle, the threshold frequency is of the order of $\,1$ yr$^{-1}\,$. Its value, though, is exponentially sensitive to the temperature and therefore may be very different for exoplanets and exomoons.} The merger of the Darwin-Kaula expansion with the
 combined rheological model, worked out in {\it{Ibid.}}, has been used to predict spin-orbit resonances of a Mercury analogue having a
 constant eccentricity and a zero obliquity (Makarov 2012) and tidal properties of super-Earths (Efroimsky 2012$\,$b).

 \subsection{Expression for the tidal torque}

 It is explained in Appendix A that the average polar component of the tidal torque can be approximated with the following expression,
 provided that (1) the obliquity is small, and (2) the perturbed body and the perturber are not too close to one another (so only the
 terms with degree-2 Love number are important)$\,$:
 \begin{subequations}
 \ba
 \langle\,{\cal{T}}_z\rangle_{\textstyle{_{\textstyle_{\textstyle{_{l=2}}}}}}~=~
 \frac{3}{2}~\frac{\,G\,M_{1}^{\,2}}{a}\,\left(\frac{R}{a}\right)^{5}\sum_{q=-1}^{7}\,G^{\,2}_{\textstyle{_{\textstyle{_{20\mbox{\it{q}}}}}}}(e)~k_2(
 \omega_{\textstyle{_{\textstyle{_{220\mbox{\it{q}}}}}}})~\sin\epsilon_2(\omega_{\textstyle{_{\textstyle{_{220\mbox{\it{q}}}}}}})\,+O(e^8\,\epsilon)+
 O(i^2\,\epsilon)~=~\quad~\quad\quad
 \label{a}
 \ea
 \ba
 \frac{3}{2}~\frac{\,G\,M_{1}^{\,2}}{a}\,\left(\frac{R}{a}\right)^{5}\sum_{q=-1}^{7}\,G^{\,2}_{\textstyle{_{\textstyle{_{20\mbox{\it{q}}}}}}}(e)~k_2(
 \omega_{\textstyle{_{\textstyle{_{220\mbox{\it{q}}}}}}})~\sin|\,\epsilon_2(\omega_{\textstyle{_{\textstyle{_{220\mbox{\it{q}}}}}}})\,|
 \,~\mbox{Sgn}\,\left(\,\omega_{220q}\,\right)+O(e^8\,\epsilon)+O(i^2\,\epsilon)~~.~\quad~
 \label{b}
 \ea
  \label{sec.eq}
 \end{subequations}
 This is the polar (orthogonal to the equator) component of the torque wherewith the tidally-perturbed body is acted upon by the perturber.
 The angular brackets denote orbital averaging, $\,G\,$ stands for the Newton gravitational constant, $\,M_{1}\,$ signifies the
 mass of the perturber (the star, if the perturbed body is its planet; or the planet, if the perturbed body is a satellite), $\,a\,$
 is the semimajor axis, while $\,R\,$ is the radius of the tidally perturbed body. The degree-2 dynamical Love number $\,k_2\,$
 and the phase lag $~\epsilon_2\,$ are functions of the Fourier tidal mode
 \ba
 \label{eco}
 \omega_{\textstyle{_{\textstyle{_{220\mbox{\it{q}}}}}}}\,=~(2~+~q)~n~-~2~\dot{\theta}\,~.
 \ea
 While the dynamical Love numbers $\,k_2(\omega_{\textstyle{_{220\mbox{\it{q}}}}})\,$ are positive definite, the sign of each phase lag $\,\epsilon_2
 (\omega_{\textstyle{_{220\mbox{\it{q}}}}})\,$ coincides with that of the Fourier mode $\,\omega_{\textstyle{_{220\mbox{\it{q}}}}}\,$, as can be
 understood from formulae (\ref{lags}) and (\ref{lags_2}) in Appendix A. It is for this reason that the products $\,k_2(\omega_{\textstyle{_{220\mbox{\it{q}}}}})\,\sin\epsilon_2(\omega_{\textstyle{_{220\mbox{\it{q}}}}})\,$ emerging in expression (\ref{a}) are
 rewritten in expression (\ref{b}) as $\,k_2(\omega_{\textstyle{_{220\mbox{\it{q}}}}})\,\sin|\,\epsilon_2
  (\omega_{\textstyle{_{220\mbox{\it{q}}}}})\,|~\,\mbox{Sgn}\,(\omega_{\textstyle{_{220\mbox{\it{q}}}}})\,$.

 A generic expression for the torque implies summation over the four integer indices $~lmpq~$ serving to number the Fourier tidal modes
 $~\omega_{{\textstyle{_{lmpq}}}}~$ entering the spectrum -- see Appendix A below. The terms of that series depend also upon the
 obliquity. As the $~lmpq~$ term contains a factor $\,(R/a)^{2l+1}\,$, it is often sufficient to keep only the degree-2 terms ($\,l=2\,$).
 In this case, with an extra assumption of small obliquity, it is enough to limit the summation to the terms with $\,m=2\,$, $\,p=0\,$.
 This renders expression (\ref{sec.eq}).

 While the full expression for the torque implies summation over all integer values of $\,q\,$, numerical tests demonstrate that, for
 eccentricities not exceeding $\,\sim\,0.3\,$, it is enough to take into account the terms up to $\,e^7\,$, inclusive. This would require
 summation from $\,q\,=\,-\,7\,$ through $\,q\,=\,7\,$. However, the values of the numerical factors entering the eccentricity polynomials
 $\,G_{\textstyle{_{\textstyle{_{20\mbox{\it{q}}}}}}}(e)\,$ are such that in practice it turns out to be sufficient to include only the
 terms with $\,q\,$ varying from $\,-\,1\,$ through $\,7\,$.

 \subsection{The tidal torque and the equilibrium eccentricity as functions of the spin rate}

 Expression (\ref{eco}) makes each product $\,k_2\,\sin\epsilon_2\,$ a function of the planetary spin rate $\,\stackrel{\bf\centerdot}{\theta~}\,$:
 \ba
 \nonumber
 k_2(\omega_{\textstyle{_{220\mbox{\it{q}}}}})\;\sin|\,\epsilon_2(\omega_{\textstyle{_{220\mbox{\it{q}}}}})\,|~\,\mbox{Sgn}\,(\,\omega_{
 \textstyle{_{220\mbox{\it{q}}}}}\,)~\quad\quad\quad\quad\quad\quad\quad\quad~\quad\quad\quad\quad\quad\quad\quad\quad\quad\quad\quad\quad
 ~\\  \nonumber\\
 =~k_2(\,2(n-\dot{\theta})\,+\,q\,n\,)~~\sin|\,\epsilon_2(\,2(n-\dot{\theta})\,+\,q\,n\,)\,|~~\mbox{Sgn}\,(\,2(n-\dot{\theta})
 \,+\,q\,n\,)\,~.~\quad
 \label{formula}
 \ea
 Consequently, the entire sum (\ref{sec.eq}) can be treated as a function of $\,\stackrel{\bf\centerdot}{\theta\,}$. The mean motion and eccentricity will play the role of
 parameters whose evolution is much slower than that of $\,\stackrel{\bf\centerdot}{\theta\,}$.
 \begin{figure}[htbp]
 \centering
 \includegraphics[angle=0,width=0.95\textwidth]{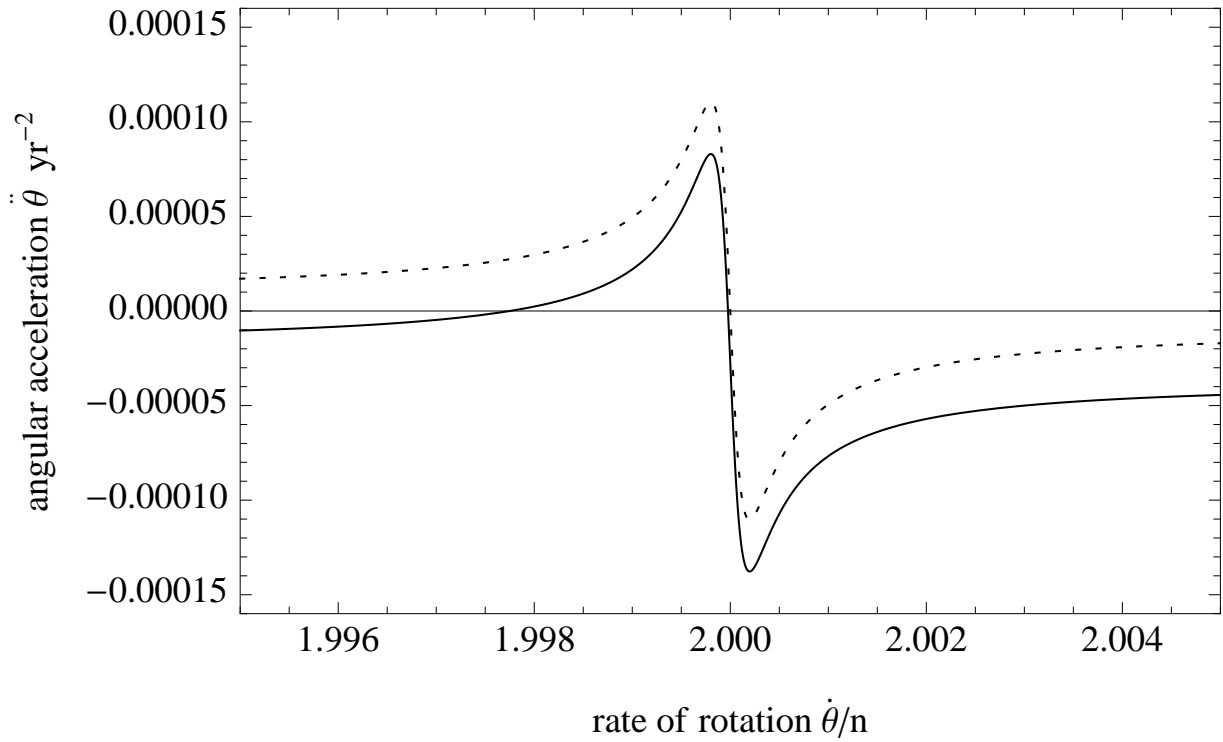}
 \caption{\small{Angular acceleration due to the secular tidal torque (\ref{sec.eq}), in the vicinity of the
 2:1 resonance.
 The dotted kink is the $\,q=2\,$ term which is an odd function when centered at $\dot{\theta}/n=1+q/2=2$. The solid line renders the total torque
 (\ref{sec.eq}), i.e.,  a sum of the $q=2$ kink and the bias comprised by the terms with $q\neq 2$. Near the resonance,
 the bias is a slowly changing function of $\dot{\theta}/n$ and can be approximated
 with a constant. The $q=2$ kink resides on the right slope of a more powerful $q=1$ kink which is centered at $\dot{\theta}/n=1$ and dominates the bias. So the $q=2$ kink is shifted downward and goes through nil a tiny bit to the left
 of the resonance.
 \vspace{1mm}
 $\quad\quad\quad\quad\quad\quad\quad\quad\quad\quad\quad\quad\quad\quad\quad\quad\quad$
 \vspace{1mm}
 The figure is borrowed from our work Makarov et al. (2012) devoted to the super-Earth GJ581d.
 }\label{tide.fig}}
 \end{figure}

 Each product $\,k_2\,\sin\epsilon_2\,$ is an odd function of the tidal mode $\,\omega_{220\mbox{\it{q}}}\,$ and has the shape of a kink centered around
 $\,\omega_{220\mbox{\it{q}}}=0\,$. When we employ relation (\ref{formula}) to write these products as functions of the spin rate, the new functions will
 still be kinks, though centered around $\,\dot{\theta}=n\left(1+{\textstyle q}/{\textstyle 2}\right)\,$. In Figure \ref{tide.fig}, the dotted line depicts
 the product \footnote{~In formula (\ref{gdwe}), the notations $~k_2(\,4\,n\,-\,2\,\dot{\theta}\,)\,$ and $~\epsilon_2(\,4\,n\,-\,2\,\dot{\theta}\,)\,$ stand for
 $\,k_2\,$ and $\,\epsilon_2\,$ as functions of the argument $\,4\,n\,-\,2\,\dot{\theta}\,$. These are {\it{not}} products of $\,k_2\,$ or $\,\epsilon_2\,$ by $\,(4\,n\,-\,2\,\dot{\theta})\,$. ~~The same pertains to formula (\ref{formula}).}
 \ba
 \nonumber
 k_2(\omega_{\textstyle{_{
 \textstyle{_{2202}}}}})~\sin\epsilon_2(\omega_{\textstyle{_{\textstyle{_{2202}}}}})\,=\,
 k_2(\,4\,n\,-\,2\,\dot{\theta}\,)~\sin\epsilon_2(\,4\,n\,-\,2\,\dot{\theta}\,)\\
 \nonumber\\
 =~k_2(\,4\,n\,-\,2\,\dot{\theta}\,)~~\sin|\,\epsilon_2(\,4\,n\,-\,2\,\dot{\theta}\,)\,|~~\mbox{Sgn}\,(\,4\,n\,-\,2\,\dot{\theta}\,)\,~.
 \label{gdwe}
 \ea
 The kink shape of the $\,k_2\,\sin\epsilon_2\,$ products is determined by the rheological properties of the planet and its self-gravitation
 (see Appendix B for details and references).
 A kink transcends nil and changes sign continuously as the spin rate goes through the appropriate resonance.

 In the sum (\ref{sec.eq}), the kink-shaped products stand with multipliers $\,G^{\,2}_{\textstyle{_{\textstyle{_{20\mbox{\it{q}}}}}}}(e)\,$. So the overall
 tidal torque (\ref{sec.eq}), as a function of $\,\dot{\theta}~$, is a superposition of many kinks having different magnitudes and centered at different resonances (nine kinks, if we sum over $\,q\,=\,-\,1,\,.\,.\,.\,7\,$). The ensuing curve will cross the horizontal axis in points extremely close to the resonances $\,\stackrel{\bf\centerdot}{\theta~}=n\,\left(1+\,{\textstyle q}/{\textstyle 2}\right)\,$, but not exactly in these resonances --- like the solid line in Figure \ref{tide.fig}.

 \subsection{The physical meaning of the kink}

 The physical forcing frequencies in the mantle, $\,\chi_{\textstyle{_{220\mbox{\it{q}}}}}\,$, are absolute values of the Fourier modes:
 \ba
 \chi_{\textstyle{_{220\mbox{\it{q}}}}}\,=~|\,\omega_{\textstyle{_{220\mbox{\it{q}}}}}\,|\,~.
 \label{}
 \ea
 The dynamical Love number is an even function of the tidal mode $\,\omega_{\textstyle{_{220\mbox{\it{q}}}}}\,$, while the phase lag is an odd function.
 For this reason, the product $\,k_2(\omega_{\textstyle{_{220\mbox{\it{q}}}}})\,\sin\epsilon_l(\omega_{\textstyle{_{220\mbox{\it{q}}}}})\,$ can be
 rewritten as a function of the physical frequency $\,\chi\,$, multiplied by the sign of the appropriate Fourier mode:
 \ba
 k_2(\omega_{\textstyle{_{220\mbox{\it{q}}}}})\,\sin\epsilon_l(\omega_{\textstyle{_{220\mbox{\it{q}}}}})\,=~k_2(\chi_{\textstyle{_{220\mbox{\it{q}}}}})\,\sin
 |\epsilon_2(\chi_{\textstyle{_{220\mbox{\it{q}}}}})|\,{\mbox{Sgn}}\,(\omega_{\textstyle{_{220
 \mbox{\it{q}}}}})\,~.
 \label{}
 \ea
 Outside the inter-peak interval
 (i.e., at frequencies that are not too low), the positive definite quantity \footnote{~This quantity is often denoted as $\,k_2/Q\,$, though notation $\,k_2\
 Q_2\,$ would be more appropriate. Tidal quality factors are not identical to the seismic quality factor, the difference becoming crucial at low frequencies.}
 $~k_2(\chi)\,\sin|\epsilon_2(\chi)|\,$ is decreasing monotonically with increase of the frequency $\,\chi=\chi_{\textstyle{_{220\mbox{\it{q}}}}}\,$. This happens
 for two reasons. One, intuitively
 obvious, is that the dynamical Love number decreases at higher frequencies, because materials are inertial, and it is getting harder for their shape
 to keep up with the varying stress as the frequency goes up. Less obvious is the circumstance that the sine of the phase lag (i.e., the inverse tidal quality
 factor), too, decreases as the frequency grows.\footnote{~To illustrate the decrease of the Love number, imagine that we dip a spoon into a bowl of honey,
 and apply to the spoon an oscillating force of a fixed amplitude. Naturally, the amplitude of motion of the spoon will be {\it{larger}} for lower
 frequencies. Sadly, this simple example will not help us to illustrate how the phase decreases with the growth of frequency. Naively, one might expect an
 anti-phase response at high frequencies, like in the case of a damped driven harmonic oscillator. This regime would indeed be taking place, had the mantle obeyed a constant time lag law. However, real minerals behave differently, so our parallels with a viscously damped oscillator or a viscous liquid have
 their limitations.}  Supported by a mighty volume of seismological, geodetic, and laboratory data, this behaviour may look
 counterintuitive because this is not what one would expect from a viscous fluid. The fact however is that at physically interesting frequencies the mantle
 behaves itself not as a viscous or a Kelvin-Vogt body but as an Andrade body dissipation wherein obeys the law $\,\sin\epsilon\propto\chi^{\,-\,\alpha}\,$,
 with $\,\alpha\approx0.14 - 0.4\,$ for most solids (Efroimsky 2012a, 2012b).

 Finally, the steep (but still continuous) near-resonant jumps connecting the peaks of the kink are explained by the fact that at
 those locations self-gravitation ``beats" rheology ({\it{Ibid.}}).

 The kink shape of $\,k_2\,\sin\epsilon_2\,$ (generally, of $\,k_l\,\sin\epsilon_l\,$) entails somewhat counterintuitive consequences for the phase lag and the geometric lag angle. Consider the principal, semidiurnal bulge. Its phase lag and the geometric lag angle are
 \ba
 \epsilon_2 (\omega_{\textstyle{_{\textstyle{_{2200}}}}})\,=\,\omega_{\textstyle{_{2200}}}~\Delta t_2 (\omega_{\textstyle{_{\textstyle{_{2200}}}}})\,=\,2\,(n\,-\,\dot{\theta})~\Delta t_2 (\omega_{\textstyle{_{\textstyle{_{2200}}}}})\,
 \label{}
 \ea
 and
 \ba
 \delta_2(\omega_{\textstyle{_{2200}}})~=~\frac{1}{2}~|\,\epsilon_2(\omega_{\textstyle{_{2200}}})\,|\,=\,|\,n\,-\,\dot{\theta}\,|~\Delta t_2 (\omega_{\textstyle{_{\textstyle{_{2200}}}}})\,~.
 \label{}
 \ea
 Were a planet composed of a material with the time lag $\,\Delta t_2 (\omega_{\textstyle{_{\textstyle{_{2200}}}}})\,$ insensitive to the value of the principal
 tidal mode $\,\omega_{\textstyle{_{2200}}}\,=\,2\,|n\,-\,\dot{\theta}|\,$, the geometric lag angle would be larger for a higher
 value of this frequency. This indeed is what one would, intuitively, expect: the higher the difference between $\,\dot{\theta}\,$ and $\,n\,$ the larger the
 angle. However, for a realistic rheology, an increase of the $\,\delta_{\textstyle{_{2200}}}\,$ angle due to an increase in $\,2\,|n\,-\,\dot{\theta}|~$
 will take place only within an extremely close proximity of the the $\,1:1\,$ resonance. Stepping beyond the kink's peak, we shall find that an increase in
 $\,2\,|n\,-\,\dot{\theta}|~$ will be accompanied by such a decrease in the time lag $\,\Delta t_2 (\omega_{\textstyle{_{\textstyle{_{2200}}}}})\,$ that the product of these two
 quantities will, overall, be decreasing with growing frequency. So both the phase lag and the geometric lag angle will become {\it{smaller}}.

 \subsection{Instability of pseudosynchronous rotation. Physical interpretation}

 Since the mode-dependence of the products (\ref{formula}) follows from the rheological properties of the mantle and from self-gravitation of the planet,
 these products turn out to be functions not only of the tidal mode $\omega_{\textstyle{_{220\mbox{\it{q}}}}}\,$, but also of the parameters defining the
 size and rheology of the perturbed body. These parameters
 (presented in Table 1) are the body's radius $\,R\,$ and mass $\,M_2\,$, as well as its unrelaxed rigidity $\,\mu\,$, Maxwell time $\,\tau_M\,$, inelastic (Andrade) time $\,\tau_A\,$, and the Andrade parameter $\,\alpha\,$.

 For a given selection of these parameters' values and a fixed spin rate $\,\dot\theta\,$, there is a unique eccentricity
 $\,e_{\rm equ}\,$ at which $~\langle\,{\cal{T}}_z
 \rangle_{\textstyle{_{\textstyle_{\textstyle{_{l=2}}}}}}=0\,$. The
 dependence of $\,e_{\rm equ}\,$ on the relative rate of rotation $\,\dot\theta/n\,$ can be found for a grid of points by numerically
 determining the roots of $~\langle\,{\cal{T}}_z
 \rangle_{\textstyle{_{\textstyle_{\textstyle{_{l=2}}}}}}~$
 in $\,e\,$. To accomplish this, we recall that each term in Equation (\ref{sec.eq}) is a polynomial in $\,e\,$ and, therefore, so is the
 entire right-hand side of (\ref{sec.eq}). Computational search of the roots was carried out for two sets of parameters listed in
 Table 1, one representing the Moon orbiting the Earth, and the other a hypothetical super-Earth orbiting a solar analog. The choice of
 parameters is intended to represent the range of applicability of the model. The resulting dependencies of $\,e_{\rm equ}\,$ upon $\,\dot\theta/n\,$
 are presented in Figures \ref{moon.fig} and \ref{super.fig}.

 The jigsaw shape of the found dependencies $\,e_{\rm equ}(\dot\theta/n)\,$ is remarkably different from the predictions of the linear torque model. The
  curves for the Moon (Figure \ref{moon.fig}) and the super-Earth (Figure \ref{super.fig}) are monotonically descending with a rising rate of rotation,
 everywhere between the spin-orbit resonances. This has profound implications for the character of equilibrium at a zero tidal torque.
 Consider an arbitrary point on the downhill portion of the curve, e.g., the one in Figure \ref{moon.fig}, from which two short arrows of
 opposite direction are drawn. A perturbation in $\,\dot\theta\,$ away from this point, whether spinning the planet up or slowing it down,
 will cause a nonzero tidal torque acting {\it in the same direction} -- as indicated by the direction of the arrows. Thereby, the tidal
 equilibrium achieved at $\,e_{\rm equ}\,$ between the resonances is inherently unstable. Similarly, a perturbation in $\,e\,$ will make
 the planet diverge from the curve of zero torque rather than return to it. Hence, the states of zero tidal torque at non-resonant
 spin rates are transient by nature.

 In a narrow vicinity of each spin-orbit resonance, $\,e_{\rm equ}\,$ takes a rapid upward increase. Computation of the roots of Equation \ref{sec.eq} is
 numerically difficult in these areas because of a very large gradient of the curve. The slope of the segments at the resonances is positive; therefore, there
 exists a stable equilibrium just as in the case of the linear torque model. A deviation of the spin rate from a resonant value gives rise to a nonzero
 restoring torque, as indicated by a pair of inward arrows in Figure \ref{moon.fig} at the resonance 3:2.

 To obtain a physical explanation of the unstable nature of the pseudosynchronous rotation, recall two circumstances. First, each term of the sum
 (\ref{sec.eq}) has the shape of a kink. Second, each such kink, as a function of $\,\dot\theta\,$, is increasing monotonically everywhere except near the appropriate resonance, as in Figure \ref{tide.fig}. For this reason, an infinitesimal increase in $\,\dot\theta/n\,$ furnishes an infinitesimal increase in the tidal torque (not necessarily in its absolute value). This happens for an arbitrary value of $\,e\,$ and for the values of $\,\dot{\theta}/n\,$ outside
 the narrow resonances. Specifically, for $\,e=e_{equ}\,$, the torque is zero and acquires a positive value, which leads to further spin-up. The spin-up continues until $\,\dot{\theta}/n\,$ stumbles into a resonance. (Resonances are depicted with near-vertical segments of the dotted curves in Figures
 \ref{moon.fig} and \ref{super.fig}.)

 \subsection{On the choice of the values for physical parameters}

 Figures \ref{moon.fig} and \ref{super.fig} reveal, in comparison, that the structure of the equilibrium tidal torque is similar for a wide range of values
 of planetary parameters. The values employed to build these plots are shown in Table \ref{table}. The rapid jumps of $\,e_{\rm equ}\,$ at resonances
 and the smooth descents between the resonances are
 characteristic of small moons and large planets likewise. The values of $\,e_{\rm equ}\,$ at resonances appear to be the same for the
 Moon and the model super-Earth. The most noticeable difference is in the amplitude of the resonance jumps, which is significantly higher
 for the Moon. By experimenting with the input parameters, we found out that this amplitude is sensitive mainly to the Maxwell time $\,
 \tau_M\,$, which differs by an order of magnitude between our model bodies.

 The choice of such a small Maxwell time for the Moon, only 5 years, is justified by the likely presence of a high percentage of partial
 melt in the lower lunar mantle. The presence of partial melt follows from the modeling carried out by Weber et al. (2011) and also from
 an earlier study by Nakamura et al. (1974). There exists data pointing at the possibility of the lunar Maxwell time being of the order
 of months. \footnote{~The smallness of the lunar $\,\tau_M\,$ ensues from the unexpected frequency dependence of the lunar tidal $\,Q\,$
 factor. According to Williams et al. (2008), the tidal $\,Q\,$ increases from $\,\sim 29\,$ at a month to $\,\sim 35\,$ at one year, a
 slope incompatible with the seismic properties of rocks which are expected to have a {\it{seismic}} $\,Q\,$ decreasing with increase of
 the period. As explained in Efroimsky (2012a), this unexpected slope may have emerged due to the difference between the tidal and
 seismic $\,Q\,$ at low frequencies. It is possible that the frequency range in which the lunar tides were studied could be close to or
 slightly left of the peak of the function $~k_2(\omega_{2200})\,\sin|\,\epsilon_2(\omega_{2200})\,|~$. In this case, $\,\tau_M\,$ of
 the Moon may be of the order of days. Fortunately, the choice of the value of $\,\tau_M\,$ does not influence considerably the jagged shape
 of the dependency $\,e_{\rm equ}(\dot{\theta}/n)\,$. The resulting plot turns out to be similar to the jagged plot for
 $\,\tau_M = 5\,$ yr depicted in Figure \ref{moon.fig}.}

 The choice of the parameters of the super-Earth was consistent with that made in Makarov et al. (2012) for the planet GJ581d.

 The Arrhenius law requires that planets and moons with hotter interiors have lower viscosity of mantles and, thus, have shorter Maxwell times. So we surmise that the secular tidal torque for such objects should be relatively more efficient in capturing at higher spin-orbit resonances.

\section{Resonant rotation of axisymmetric bodies}

\begin{figure}[htbp]
  \centering
  \subfigure[]
  {
    \includegraphics[width=77mm]{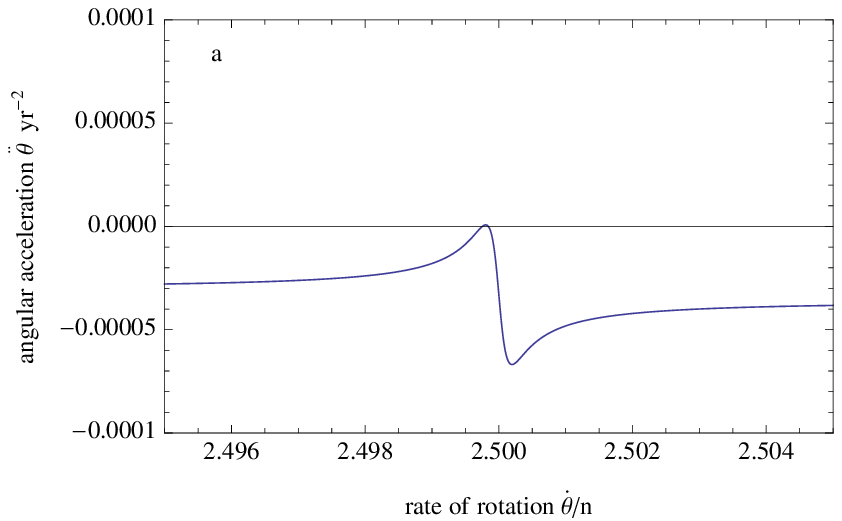}
    }
    \subfigure[]
    {
    \includegraphics[width=85mm]{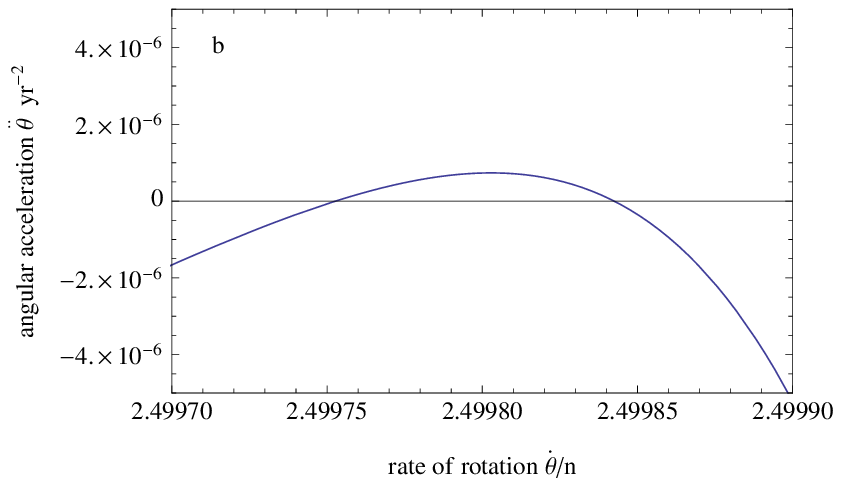}
    }
 \caption{\small{Angular acceleration of the super-earth (Table 1) caused by the secular tidal torque
in the vicinity of the 5:2 resonance, (a) showing the entire resonance interval, and (b) showing in
more detail the same curve in the area of the local maximum.}
\label{sec.fig}}
\end{figure}

 We have determined that a stable spin-orbit equilibrium is achieved at spin rates where the value of the secular polar tidal
 torque is zero, while the derivative of the equilibrium eccentricity with respect to the spin rate, $~{\textstyle de_{equ}}/{
 \textstyle d(\dot{\theta}/n)}~$, is negative. With the realistic tidal model discussed in Appendix B, this may happen only in the
 vicinity of spin-orbit resonances because the derivative of the torque is positive elsewhere. The secular torque (\ref{sec.eq}), as well as
 the angular acceleration caused by it, has ``kinks" in the vicinity of spin-orbit commensurabilities $\,\dot\theta/n\,=\,1\,+\,q/2~$,
 with an integer $\,q\,$ of either sign. An example thereof is shown in Figure \ref{sec.fig} for the super-Earth model, within a segment
 of the spin rate around $\,\dot\theta/n=5/2\,$. The kink is comprised by a local maximum below the resonance and a minimum above the
 resonance.

 To understand the plot in Figure \ref{sec.fig}, recall that in the vicinity of each resonance $\,q
 \,'\,$, i.e., for $\,\dot{\theta}/n\,$ being close to $~1\,+\,{\textstyle q\,'}/{\textstyle 2}~$, the right-hand side of (\ref{sec.eq})
 can be decomposed into two parts.
 One part is the $\,q=q\,'\,$ term. It is a kink-looking odd function of the tidal mode $\,\omega_{220q}\,$, and it goes through nil at
 exactly $\,\omega_{220q}=0~$. Due to (\ref{eco}), this term can also be interpreted as a function of the spin rate, antisymmetric
 with respect to the resonance point $\,\dot{\theta}/n\,=\,1\,+\,{\textstyle q\,'}/{\textstyle 2}\,$ where this term goes through nil.
 The second part, called {\it{bias}}, is the rest of the sum, i.e., the input of all the $\,q\neq q\,'\,$ Fourier modes into the values
 assumed by the torque in the vicinity of the $\,q=q\,'\,$ resonance. The bias can be negative or positive in value, depending on the
 eccentricity. For not too large eccentricities, it is usually negative. Being a very slowly changing function within the resonance
 interval, it can, to a good approximation, be assumed constant there.

 Having summed up all the terms in (\ref{sec.eq}), and exploring the behaviour of this sum near $\,\dot{\theta}/n\,=\,1\,+\,{\textstyle q
 \,'}/{\textstyle 2}\,$, we see that the resulting curve does {\it{not}} cross the horizontal axis at $\,\dot{\theta}/n\,=\,1\,+\,
 {\textstyle q\,'}/{\textstyle 2}\,$. One can say that the bias slightly displaces the location of zeros. The zeros are located close to
 resonances but not exactly in resonances.

 In Figure \ref{sec.fig}(a), the values of the overall torque (in fact, of the total angular acceleration proportional to the torque) are
 defined mostly by the $\,q=3\,$ term which, in this vicinity, looks like an antisymmetric kink. However, the curve is shifted down due to the bias
 which is defined mainly by the right slope of the $\,q=1\,$ kink located to the left. Since the right slope of the $\,q=1\,$ kink is
 negative, the $\,q=3\,$ kink is shifted down. As a result, the maximum secular torque barely rises above zero, and the curve has two
 zeros located to the left of the point $\,\dot{\theta}/n\,=\,5/2\,$, close to the maximum of the kink. The interval of the
 resonance is defined by the location of the peaks: $\,\dot\theta/n=2.4998\,$ and $\,2.5002\,$.

 Figure \ref{sec.fig}(b) is a blow-up of Figure \ref{sec.fig}(a) showing in greater detail the area around the maximum
 of the kink. The root of the function within the resonance interval is at $\,2.49985\,$ rather than exactly $\,2.5\,$. For the same
 reason, the torque value is negative at $\,\dot\theta/n=2.5\,$.

 In the framework of this model, it is reasonable to define capture in resonance as an equilibrium state in which the body's average
 rotation rate stays between the values corresponding to these two peaks. This definition is adequate because, as we saw above, stable
 equilibrium is possible only on negative slopes of the angular acceleration (or tidal torque) as a function of the spin rate.

\begin{figure}
\includegraphics[width=95mm]{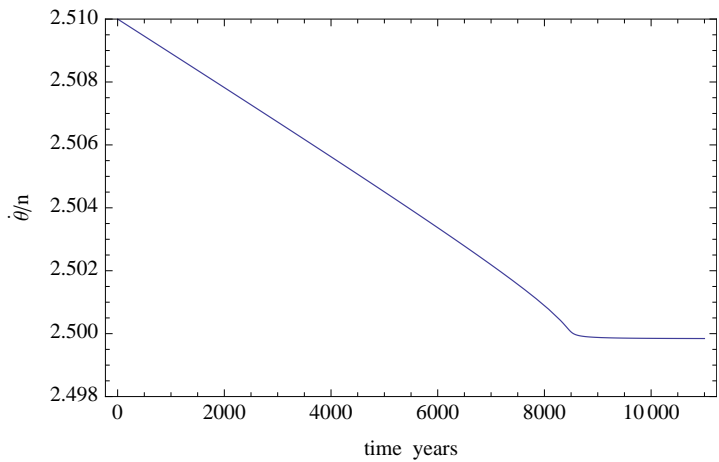}
\caption{\small{Capture of an axisymmetric super-Earth (Table 1, but with $(B-A)/C=0$) in 5:2 resonance.
Note that the ultimate equilibrium spin rate is slightly less than $2.5\,n$.}
\label{capt.fig}}
\end{figure}

 When a triaxial planet is captured in a 5:2 resonance, its time-averaged spin rate is exactly $\,2.5\,n\,$. Similarly, the Moon's spin
 rate, captured in synchronous rotation, is exactly $\,1\,n\,$. However, we know that the time-averaged tidal torque is nonzero at
 this spin rate. Why does not the Moon accelerate? For a triaxial body, the nonzero secular tidal torque is compensated by a
 counteracting triaxiality-caused torque, through a small tilt of the average inertia axis with respect to the line connecting the
 centres of the bodies. A nonzero time-average tilt generates a secular triaxial torque. This mechanism of torque compensation obviously
 does not work for the rather hypothetical case of axisymmetric body, which would be subject to only tidal forces. Would the Moon be
 facing the Earth always with the same side if it were completely axisymmetric? First, we have to find out if capture in spin-orbit
 resonance is at all possible. The answer is yes, as long as the secular torque changes sign in the vicinity of that resonance
 \footnote{~Triaxial bodies can be captured in spin-orbit resonances even if the secular tidal torque is negative everywhere in the
 vicinity of that resonance.}. Even though the maximum torque in Figure \ref{sec.fig} barely rises above zero at spin rates below the
 resonance, it turns out to be sufficient for capture in 5:2. Figure \ref{capt.fig} displays the results of a numerical integration
 of the evolution of spin rate for the super-Earth model (Table 1) at $\,\tau_M=50$ yr with an initial spin rate of $\,\dot\theta(0)=
 2.51\,n\,$. The planet is captured in about 8500 yr, but the equilibrium spin rate at $\,2.49985\,n\,$ is clearly below the resonance
 value. This value is consistent with the root of the secular torque within the resonance interval, Figure \ref{sec.fig}(b). Thus, the
 equilibrium resonance state of an axisymmetric body is achieved at the spin rate where the secular tidal torque equals zero, as expected.

 \section{Word of caution}

 As was demonstrated above, stability of pseudosynchronous spin hinges upon rheology. Being
 stable for the constant time lag model, the regime is expected to be transient for realistic mantles, insofar as their $\,k_2(\chi)\,\sin\epsilon_{2}(\chi)\,$ has one
 pronounced peak (not to count the opposite one at the negative value of the tidal mode) -- see Figure \ref{tide.fig}. The situation will have to be
 re-analysed for bodies of complex structure (with surface or internal oceans), as well as for bodies of yet unexplored rheologies. Specifically, if it
 happens that somewhere in universe there exist bodies with not one but two pronounced peaks of $\,k_2\,\sin\epsilon_2\,$ at positive frequencies, the
 ``ditch" between these peaks may, in principle, lead to emergence of a pseudosynchronous rotation state. Such a peak may emerge at the boundary of two
 frequency bands dominated by different friction mechanisms, i.e., when a new mechanism is ``turned on" very quickly with the increase of frequency.
 Although highly hypothetical, such situations should not be written off completely.

 \section{Discussion}

 The shortness of Earth's day was undoubtedly beneficial for proliferation of biological life,  making the daily temperature
 variation moderate. The situation may be drastically different for the  growing class of potentially habitable super-Earth exoplanets.
 Due to the observational selection  effect, the spectroscopically detected super-Earths are found mostly around lower-mass stars, whose
 habitable zones are inevitably narrower and closer. For such systems, any conclusion about potential habitability of a given exoplanet becomes especially
 uncertain, and the analysis becomes intricately involved with regard to such parameters as the amount of stellar irradiation, the
 chemical composition of a hypothetical atmosphere, and the internal heating. The rate of rotation is also a crucial parameter which for
 now remains unavailable from observation. The most advanced climatic simulations are based on a certain assumption of the spin-orbit
 state of the planet, e.g., a tidal synchronisation (1:1 resonance) is assumed, as in Selsis et al (2011). A tidally synchronised planet
 showing the same side to its host star has this side always exposed to plentiful irradiation, the other side staying dark.
 Such planets can hardly retain an atmosphere and can hardly be habitable. However, a spin-orbit locking into higher commensurabilities
 (e.g., 3:2, as in the case of Mercury) allows the planet to rotate with respect to the  host star, and leaves the planet a possibility
 of sustaining a stable atmosphere and having water in the liquid form on the surface. Three-dimensional climatic simulations of the potentially
 habitable super-Earth GJ 581d, by Wordsworth et al. (2010), were performed for a set of possible spin-orbit resonances, including 2:1.
 As was later explained in Makarov et al. (2012), this resonance is the likeliest state of GJ 581d for a wide range of rheological parameters, assuming a terrestrial composition of its mantle. Wordsworth et al. (2010) drew attention to the fact that a tidally synchronised
 atmosphere may be short-lived because of the collapse of CO$_2$ on the night side. A similar phenomenon may occur on a slowly rotating
 planet. Both the constant angular lag tidal model and the constant time lag model predict that oblate planets with moderate and large
 eccentricities are captured in stable pseudosynchronous rotations, in which case their spin rate only slightly exceeds their orbital
 mean motion. This would make an entire class of detected exoplanets unsuitable for biological life. In this paper we prove that the
 prediction of pseudo-synchronism is germane to the above-mentioned simplified models of tidal interactions, models inapplicable to
 solid planets or moons. Super-Earth exoplanets of Earth-like composition on eccentric orbits are likely to be captured into spin-orbit
 resonances higher than 1:1, but there is no such thing as pseudo-synchronous rotation for these objects.

 \section*{Acknowledgments}

 The authors would like to express their thanks to the referee, Stanton Peale, who provided several incisive comments and criticisms, and who
 urged the authors to present physical interpretation of the results obtained in the paper. One of the authors (ME) is indebted to Sylvio Ferraz Mello
 and James G. Williams for numerous stimulating discussions on the topic of this work.

 This research has made use of NASA's Astrophysics Data System.

 \section*{\underline{\textbf{\Large{Appendix $\,$A.}}}
 ~~~~~\Large{The tidal torque~~~~~~~~~~~~~~~~}}

 The additional potential $\,U\,$ of a tidally perturbed body can be expanded into a Fourier series over the tidal modes
 \ba
 \omega_{lmpq}\;\equiv\;({\it l}-2p)\;\dot{\omega}\,+\,({\it l}-2p+q)\;{\bf{\dot{\cal{M}}}}\,+\,m\;(\dot{\Omega}\,-\,\dot{\theta})
 \,\approx\,(l-2p+q)\,n\,-\,m\,\dot{\theta}\,~,~~~
 \label{L9}
 \ea
 where $\,{\theta}\,$ and $\,\dot{\theta}\,$ are the sidereal angle and rotation rate of the body, while $\,\omega\,$,
 $\,\Omega\,$, $\,n\,$, and $\,{\cal{M}}\,$ are the periapse, the node, the mean motion, and the mean anomaly of the perturber as seen
 from the perturbed body.

 While the tidal modes $\,\omega_{\textstyle{_{lmpq}}}\,$ can be of either sign, the forcing frequencies
 \ba
 \chi_{lmpq}\,=~|\,\omega_{lmpq}\,|~\approx~|~(l-2p+q)~n\,-\,m~\dot{\theta}~|
 \label{fr}
 \ea
 at which the strain and stress oscillate are positive-definite.

 The series expansion of the additional potential $\,U\,$ was developed by Kaula (1964), its partial sum known yet to Darwin (1879).
 Therefore the series for $\,U\,$ and the resulting series for the torque are often named {\it{the Darwin-Kaula expansions}}.

 An accurate derivation of the expansion for the torque demonstrates that the torque contains both a rapidly oscillating and a secular
 part (Efroimsky 2012a). Having a zero orbital average, the oscillating part nevertheless may play a role in dissipation of free
 librations. In Makarov et al. (2012) it was explored whether the oscillating part of the torque can influence capture into resonances.
 Changing the outcome of a particular realisation of the capture scenario, the oscillating part did not alter the statistics.

 The secular part of the polar torque looks as
 \ba
 \langle\,{\cal{T}}_z
 \rangle\,=\,2\,G\,M_{star}^{{{\,2}}}
 \sum_{{\it{l}}=2}^{\infty}
 \frac{R^{\textstyle{^{2l\,+\,1}}}}{
 a^{\textstyle{^{2l\,+\,2}}}}
 \sum_{m=0}^{l}
 \frac{(l-m)!}{(l+m)!}\;m
 \sum_{p=0}^{l}F^{\,2}_{lmp}(i)\sum^{\it \infty}_{q=-\infty}
 G^{\,2}_{lpq}(e)\;k_l(\omega_{lmpq})\;\sin\epsilon_l(\omega_{lmpq})\,~,\quad
 \label{31}
 \ea
 where the angular brackets signify orbital averaging, $\,G\,$ denotes Newton's gravity constant, $\,a,\,i,\,e\,$ are the semimajor
 axis, inclination (or obliquity), and eccentricity, while $\,F_{lmp}(i)\,$ and $\,G_{lpq}(e)\,$ are the inclination functions and eccentricity
 polynomials. The Love numbers $\,k_{\textstyle{_l}}\,$ and the phase lags $\,\epsilon_{\textstyle{_l}}\,$ depend on the modes (\ref{L9}).

 In the Darwin-Kaula theory, the phase lags come into being as products of the modes $\,\omega_{lmpq}\,$ by the corresponding time delays:
 \ba
 \epsilon_l(\omega_{lmpq})\,=\,\omega_{lmpq}~\,\Delta t_l(\omega_{lmpq})\,~,
 \label{lags}
 \ea
 where, for causality reasons, the time lags $\,\Delta t_{\textstyle{_l}}(\omega_{\textstyle{_{lmpq}}})\,$ are positive-definite. Therefore,
 (\ref{lags}) may be rewritten as
 \ba
 \epsilon_l(\omega_{lmpq})\,=\,\chi_{lmpq}~\,\Delta t_l(\omega_{lmpq})~\,\mbox{Sgn}\,(\,\omega_{lmpq}\,)\,~,
 \label{lags_2}
 \ea
 $\chi_{lmpq}\,$ being the physical forcing frequencies (\ref{fr}). As a result of this, the entire expression for the polar component of
 the torque can be written down as
 \ba
 \nonumber
 \langle\,{\cal{T}}_z
 \rangle\,=
 ~~~\quad~\quad~\quad~\quad~~\quad~\quad~\quad~\quad~\quad~\quad~\quad~\quad~\quad~\quad~\quad~\quad~\quad~\quad~\quad~\quad~\quad~\quad~\quad~\quad~\quad~\quad~\quad~\quad
 \\
 \nonumber\\
 2\,GM_{star}^{{{\,2}}}
 \sum_{{\it{l}}=2}^{\infty}
 \frac{R^{\textstyle{^{2l\,+\,1}}}}{
 a^{\textstyle{^{2l\,+\,2}}}}
 \sum_{m=0}^{l}
 \frac{(l-m)!}{(l+m)!}\;m
 \sum_{p=0}^{l}F^{\,2}_{lmp}(i)\sum^{\it \infty}_{q=-\infty}
 G^{\,2}_{lpq}(e)\;k_l(\omega_{lmpq})~\sin|\,\epsilon_l(\omega_{lmpq})\,|\,~\mbox{Sgn}\,\left(\,\omega_{lmpq}\,\right)
 \,~.\quad
 \label{3111}
 \ea
  Be mindful that, similar to the Love numbers, we prefer to denote the lags with
 \footnote{~Although Kaula (1964) denoted the phase lags with $\,\epsilon_{{\textstyle{_{lmpq}}}}\,$, the notation $\,\epsilon_l
 (\omega_{{\textstyle{_{lmpq}}}})\,$ is more logical. It serves to emphasise the fact that for a homogeneous near-spherical body the
 functional dependence of a lag upon the tidal mode is defined by the degree $\,l\,$ solely, while the dependence of the lag upon
 $\,m,\,p,\,q\,$ comes about only due to the tidal mode $~\omega_{{\textstyle{_{lmpq}}}}~$ being dependent on these integers.

 While in the case of triaxial bodies the functional form of the lags is parameterised by all the four integers, for small triaxiality
 this complication may be ignored.}
  $\,\epsilon_l(\omega_{lmpq})\,$ and $\,\Delta t_l(\omega_{lmpq})\,$, and not with $\,\epsilon_{lmpq}\,$ and $\,\Delta t_{lmpq}\,$.

 When the bodies are not too close (${\textstyle R}/{\textstyle a}\ll1$), we drop the terms with $\,l > 2\,$. For small obliquities
 ($i\simeq 0$), we leave only $\,i-$independent terms (the next-order terms being quadratic in $\,i\,$). Finally, for eccentricities not
 exceeding $\,\sim\,0.3\,$, only the terms up to $\,e^7\,$ are important. Formally, this would imply summation over $\,q\,=\,-\,7,\,.\,.\,.
 7\,$. However, the term with $\,q\,=\,-\,2\,$ vanishes identically, while those with $\,q\,=\,-\,7,\,.\,.\,.\,-\,3\,$ are accompanied
 with extremely small numerical factors and can thus be dropped. So the polar component of the torque is approximated with
 \ba
 \nonumber
 \langle\,{\cal{T}}_z
 \rangle_{\textstyle{_{\textstyle_{\textstyle{_{l=2}}}}}}~=~~\quad~\quad~~\quad~\quad~\quad~\quad
 ~\quad~\quad~\quad~\quad~\quad~\quad~\quad~\quad~\quad~\quad~\quad~\quad~\quad~\quad~\quad~\quad~\quad~\quad~\quad~\quad\\
 \nonumber\\
 \frac{3}{2}~G\,M_{star}^{\,2}\,R^5\,a^{-6}\sum_{q=-1}^{7}\,G^{\,2}_{\textstyle{_{\textstyle{_{20\mbox{\it{q}}}}}}}(e)~k_2(
 \omega_{\textstyle{_{\textstyle{_{220\mbox{\it{q}}}}}}})~\sin|\,\epsilon_2(\omega_{\textstyle{_{\textstyle{_{220\mbox{\it{q}}}}}}})\,|
 \,~\mbox{Sgn}\,\left(\,\omega_{220q}\,\right)+O(e^8\,\epsilon)+O(i^2\,\epsilon)~~,~\quad~\quad~
 \label{16b}
 \ea
 For the first time, this expression (with a sum running over all integer $\,q\,=\,-\,\infty,\,.\,.\,.\,\infty\,$) was written, with no
 proof, by Goldreich \& Peale (1966). A sketch of a proof was later suggested by Dobrovolskis (2007).

 The functional form of the dependence
 of the factors $~k_l\,\sin\epsilon_l~$ upon the mode $\,\omega_{\textstyle{_{lmpq}}}\,$ is determined by the size and mass of
 the body and by its rheological properties. By rheology we imply the so-called constitutive equation of the medium, i.e., an equation
 interconnecting the strain and stress. For linear deformations, such equations can be rewritten in the frequency domain where each
 harmonic mode of the strain becomes expressed {\it{algebraically}} through the appropriate harmonic mode of the stress. Using the
 method explained in Efroimsky (2012a, 2012b), the algebraic relations can be used to find the shape of the functions $\,k_l(\omega_{
 \textstyle{_{lmpq}}})~\sin\epsilon_l(\omega_{\textstyle{_{lmpq}}})\,$ standing in the terms of the Darwin-Kaula expansion of the tidal
 torque.

 Calculations of the shapes of these functions, presented in {\it{Ibid.}}, are based on a combined rheological model (Andrade at higher
 frequencies, Maxwell at lower frequencies), because of this model's ability to best match laboratory experiments and both seismic and
 geodetic measurements of dissipation over a broad range of frequencies in the solid Earth. It is reasonable to assume that the model
 is applicable to the mantles of other terrestrial planets and moons. As demonstrated in {\it{Ibid.}}, this combined model furnishes for
 $\,k_l~\sin\epsilon_l\,$ a kink-shaped dependence upon the Fourier mode -- as in Figure \ref{sec.fig}.

 \section*{\underline{\textbf{\Large{Appendix $\,$B.}}}
 ~~~~{{\Large{Calculation of the factors}} $~k_l(\omega_{\textstyle{_{lmpq}}})~\sin\epsilon_l(\omega_{\textstyle{_{lmpq}}})~$}}

 As explained in Efroimsky (2012a, 2012b), the products $~k_l(\omega_{\textstyle{_{lmpq}}})~\sin\epsilon_l(\omega_{\textstyle{_{lmpq}}})~$ can
 be expressed through the mass and radius of the planet, and the real and imaginary parts of the complex compliance of its mantle.
 Thereby, the shape of the functional dependence of $~k_l~\sin\epsilon_l~$ upon $\,\omega_{\textstyle{_{lmpq}}}\,$ is defined by both the
 self-gravitation of the planet and its rheological properties. The functions turn out to be odd. They go continuously through nil,
 changing their sign, when the argument $\,\omega_{\textstyle{_{lmpq}}}\,$ goes through nil, i.e., when a commensurability is crossed.

 These odd functions can then be written down as $~k_l(\omega_{\textstyle{_{lmpq}}})~\sin|\,\epsilon_l(\omega_{\textstyle{_{lmpq}}})\,
 |~\,\mbox{Sgn}\,(\,\omega_{\textstyle{_{lmpq}}}\,)~$. Here the product $~k_l(\omega_{\textstyle{_{lmpq}}})~\sin|\,\epsilon_l(\omega_{
 \textstyle{_{lmpq}}})\,|~$ is an $\,${\it{even}}$\,$ function of the tidal mode and can thus be treated as a function not of the tidal
 mode $\,\omega_{\textstyle{_{lmpq}}}\,$ but of its absolute value $\,\chi_{\textstyle{_{lmpq}}}\,=\,|\,\omega_{\textstyle{_{lmpq}}}\,|\,
 $, which is the actual frequency of the oscillating stress in the mantle:
 \ba
 \nonumber
 k_l(\omega_{\textstyle{_{lmpq}}})~\sin\epsilon_l(\omega_{\textstyle{_{lmpq}}})&=&k_l(\omega_{\textstyle{_{lmpq}}})~\sin|\,\epsilon_l
 (\omega_{\textstyle{_{lmpq}}})\,|~\,\mbox{Sgn}\,(\,\omega_{\textstyle{_{lmpq}}}\,)\\
 \nonumber\\
 &=&k_l(\chi_{\textstyle{_{lmpq}}})~\sin|\,\epsilon_l
 (\chi_{\textstyle{_{lmpq}}})\,|~\,\mbox{Sgn}\,(\,\omega_{\textstyle{_{lmpq}}}\,)\,~.
 \label{qq1}
 \ea
  The development in {\it{Ibid.}} results in the following frequency dependence:
 \ba
 k_l(\omega_{\textstyle{_{lmpq}}})\;\sin\epsilon_l(\omega_{\textstyle{_{lmpq}}})\,=\;\frac{3}{2\,({\it l}\,-\,1)}\;\,\frac{-\;A_l\;J\;{\cal{I}}{\it{m}}\left[\bar{J}(\chi)\right]}{
 \left(\;{\cal{R}}{\it{e}}\left[\bar{J}(\chi)\right]\;+\;A_l\;J\;\right)^2\;+\;\left(\;{\cal{I}}{\it{m}}\left[\bar{J}(\chi)\right]\;
 \right)^2} ~\,\mbox{Sgn}\,(\,\omega_{\textstyle{_{lmpq}}}\,)~~~,~~~~~
 \label{L39b}
 \ea
 where $\,\chi\,$ is a shortened notation for the frequency $\,\chi_{\textstyle{_{lmpq}}}\,$, while coefficients $\,A_l\,$ are given by
 \ba
 A_{\it l}\,
 \equiv\;\frac{\textstyle{(2\,{\it{l}}^{\,2}\,+\,4\,{\it{l}}\,+\,3)\,{\mu}}}{\textstyle{{\it{l}}\,\mbox{g}\,
 \rho\,R}}\;=\;\frac{\textstyle{3\;(2\,{\it{l}}^{\,2}\,+\,4\,{\it{l}}\,+\,3)\,{\mu}}}{\textstyle{4\;{\it{l}}
 \,\pi\,G\,\rho^2\,R^2}}\;\;\;.~~~~~~~
 \label{qq2}
 \ea
 with $\,G\,$ being the Newton gravitational constant, and $\,R\,$, $\,\rho\,$, $\,\mu\,$, and g being the radius, mean density,
 unrelaxed rigidity, and surface gravity of the planet. The functions
 \ba
 {\cal R}{\it e} [ \bar{J}(\chi)]\;=\;J\;+\;J\,(\chi\tau_{_A})^{-\alpha}\;\cos\left(\,\frac{\alpha\,\pi}{2}\,\right)
 \;\Gamma(\alpha\,+\,1)~~~~\quad\quad\quad\quad\quad~\quad\quad\quad\quad\quad\quad\quad
 \label{A4ccc}
 \ea
 and
 \ba
 {\cal I}{\it m} [ \bar{J}(\chi)]\;=\;-\;J~(\chi\tau_{_M})^{-1}\;-\;J\,(\chi\tau_{_A})^{-\alpha}\;\sin\left(
 \,\frac{\alpha\,\pi}{2}\,\right)\;\Gamma(\alpha\,+\,1)~~~~~~~~~~~~~~~~~~\quad\quad\quad
 \label{A3ccc}
 \ea
 are the real and imaginary parts of the complex compliance $\,\bar{J}(\chi)\,$ of the mantle. Here $\,\alpha\,$ is the Andrade parameter
 assuming values of about $\,0.3\,$ for solid silicates and about $\,0.14 - 0.2\,$ for partial melts. In our computations, we used $\,
 \alpha=0.2\,$. The quantity $\,J\,$ is the unrelaxed compliance of the mantle, which is the inverse of the mantle's unrelaxed rigidity
 $\,\mu\,$. The parameters $\,\tau_{_M}\,$ and $\,\tau_{_A}\,$ are typical timescale characterising the mantle's viscoelastic and
 inelastic response, correspondingly.

 The Maxwell time $\,\tau_{_M}\,$ is the ratio of the mantle's viscosity $\,\eta\,$ to its rigidity $\,\mu\,$. While for
 the Earth's mantle it has a value of about 500 years, it may be much shorter for warmer planets and moons due to the exponential
 temperature-dependence of the viscosity.

 The inelastic (Andrade) time $\,\tau_{_A}\,$ is expected to be of the same order as or lower than $\,\tau_{_M}\,$, over the frequencies
 higher than some threshold. For these frequencies then, the inelastic (containing $\,\tau_{_A}\,$) terms in (\ref{A4ccc} - \ref{A3ccc})
 will be comparable to or larger than the viscoelastic (containing $\,\tau_{_M}\,$) terms. However, at frequencies below the threshold,
 inelasticity ceases to play a major role in the internal friction, giving way to viscous friction which becomes dominant. Therefore at
 very low frequencies the mantle's behaviour approaches that of a Maxwell body. Mathematically, this implies that below the threshold the
 parameter $\,\tau_{_A}\,$ increases rapidly as the frequency goes down (Efroimsky 2012a, 2012b). So only the first term in (\ref{A4ccc}) and
 the first term in (\ref{A3ccc}) are important, and we arrive at the complex compliance of a Maxwell material. The location of the
 frequency threshold may vary considerably for different planets. For the Earth, it is of the order of 1 yr$^{-1}\,$ (Karato \& Spetzler
 1990).

 Numerical computations show that the probabilities of capture into resonances are not very sensitive to the value of $\,\tau_{_A}\,$,
 nor to the location of the threshold, nor to how quickly inelasticity yields to viscoelasticity with the decrease of frequency. In our
 numerics, we treat $\,\tau_{_A}\,$ in the same way as in Makarov et al. (2012) and Makarov (2012). We set the threshold to be the same
 as for the solid Earth, 1 yr$^{-1}\,$. We then kept $\,\tau_{_A}\,=\,\tau_{_M}\,$ over the frequencies above the threshold. For
 frequencies lower than the threshold, we set $\,\tau_{_A}\,$ to grow exponentially with the decrease of the frequency. This way, at low
 frequencies the rheological model approaches the Maxwell one exponentially. For details, see {\it{Ibid}}.

 Writing a code, it is easier to divide both the numerator and denominator of (\ref{L39b}) by $\,J^{\,2}\,$:
 \ba
 k_l(\,\omega_{\textstyle{_{lmpq}}}\,)\;\sin\epsilon_l(\,\omega_{\textstyle{_{lmpq}}}\,)\;=\;\frac{3}{2\,({\it l}\,-\,1)}\;\,\frac{-\;A_l\;{\cal{I}}}{\left(\;{\cal{R}}
 \;+\;A_l\;\right)^2\;+\;{\cal{I}}^{\textstyle{^{\,2}}}}~\,\mbox{Sgn}\,(\,\omega_{\textstyle{_{lmpq}}}\,) ~~~,~~~~~
 \label{L39}
 \ea
 where $\,{\cal R}\,$ and $\,{\cal I}\,$ are the {\it{dimensionless}} real and imaginary parts of the
 complex compliance:
 \ba
 {\cal R}\;=\;1\;+\;(\chi\tau_{_A})^{-\alpha}\;\cos\left(\,\frac{\alpha\,\pi}{2}\,\right)
 \;\Gamma(\alpha\,+\,1)~~~,\quad\quad\quad\quad\quad~\quad\quad\quad\quad\quad\quad\quad
 \label{A4c}
 \ea
 \ba
 {\cal I}\;=\;-\;(\chi\tau_{_M})^{-1}\;-\;(\chi\tau_{_A})^{-\alpha}\;\sin\left(
 \,\frac{\alpha\,\pi}{2}\,\right)\;\Gamma(\alpha\,+\,1)~~~,~~~~~~~~~~~~~~\quad\quad\quad\quad\quad
 \label{A3c}
 \ea
 $\chi\,$ being a short notation for the physical forcing frequency $\,\chi_{\textstyle{_{lmpq}}}\,\equiv\,|\,\omega_{\textstyle{_{lmpq}}}\,|\,$.



\begin{thebibliography}{}


 \bibitem{}
 Bambusi, D., \& Haus, E. 2012. ``Asymptotic stability of synchronous orbits
 for a gravitating viscoelastic sphere." {\it{Celestial Mechanics and Dynamical Astronomy}}, Vol. {\bf{114}}, pp. 255 - 277\\
 http://arxiv.org/abs/1012.4974


 \bibitem{} Darwin, G. H. 1879. ``On the precession of a viscous spheroid
            and on the remote history of the Earth." {\it{Philosophical
            Transactions of the Royal Society of London}}, Vol.
            {\bf{170}}, pp. 447 - 530

 \bibitem{} Dobrovolskis, A. 2007. ``Spin states and climates of eccentric exoplanets." {\it{Icarus}}, Vol. {\bf{192}}, pp. 1 - 23

 \bibitem{}
 Efroimsky, M., \& Lainey, V. 2007.
 ``The Physics of Bodily Tides in Terrestrial Planets, and the Appropriate Scales of Dynamical Evolution."
 {\it{Journal of Geophysical Research -- Planets}}, Vol. {\bf{112}}, ~id. E12003. ~~~~doi:10.1029/2007JE002908\\
 http://arxiv.org/abs/0709.1995

 \bibitem{}
 Efroimsky, M., \& Williams, J. G. 2009. ``Tidal torques. A
 critical review of some techniques." {\it{Celestial Mechanics and Dynamical Astronomy}}, Vol. {\bf{104}}, pp. 257 - 289\\http://arxiv.org/abs/0803.3299

 \bibitem{}
 Efroimsky, M. 2012$\,$a.
 ``Bodily tides near spin-orbit resonances." {\it{Celestial Mechanics and Dynamical Astronomy}}, Vol. {\bf{112}}, pp. 283 - 330.
  ~~Extended version: ~http://arxiv.org/abs/1105.6086

 \bibitem{}
 Efroimsky, M. 2012$\,$b.
 ``Tidal dissipation compared to seismic dissipation: in small bodies, earths, and superearths."
 {\it{The Astrophysical Journal}}, Vol. {\bf{746}}, ~id. 150.\\
 doi:10.1088/0004-637X/746/2/150 ~~~~http://arxiv.org/abs/1105.3936\\
 ERRATA: {\it{ApJ}}, Vol. {\bf{763}}, ~id. $\,$150 (2013)

 \bibitem{}
 Efroimsky, M., and Makarov, V. V. 2013.
 ``Tidal Friction and Tidal Lagging. Applicability Limitations of a Popular Formula for the Tidal Torque." {\it{The Astrophysical Journal}}, Vol. {\bf{764}}, ~id. $\,$26\\
 http://arxiv.org/abs/1209.1615

 \bibitem{}
 Eggleton, P. P.; Kiseleva, L. G.; and Hut, P. 1998.
 ``The equilibrium tide model for tidal friction." {\it{The Astrophysical Journal}}, Vol. {\bf{499}}, pp. 853 - 870

 \bibitem{}
 Eggleton, P.P. 2011. ``The Equilibrium Tide
  and Tidal Friction", priv. comm., available from author upon request

 \bibitem{}
 Fekel, F.C.; Browning, J.C.; Henry, G.W.; Morton, M.D.; and Hall, D.S.
 1993. ``Chromospherically active stars. X - Spectroscopy and photometry of HD 212280."
 {\it{The Astronomical Journal}}, Vol. {\bf{105}}, pp. 2265 - 2275

 \bibitem{}
 Fekel, F.C., et al. 1998.
 ``Chromospherically Active Stars. XVII. The Double-lined Binary 54 Camelopardalis (AE Lyncis)."
 {\it{The Astronomical Journal}}, Vol. {\bf{115}}, pp. 1153 - 1159


 \bibitem{} Ferraz-Mello, S. 2012. ``Tidal synchronisation of close-in satellites and exoplanets. A rheophysical approach." Submitted to {\it{Celestial
 Mechanics and Dynamical Astronomy}}\\ ~http://arxiv.org/abs/1204.3957


 \bibitem{}
 Garaud, P.; Ogilvie, G. I.; Miller, N.; and Stellmach, S. 2010.
 ``A model of the entropy flux and Reynolds stress in turbulent convection."
 {\it{Monthly Notices of the Royal Astronomical Society of London}}, Vol. {\bf{407}}, pp. 2451 - 2467\\
 http://arxiv.org/abs/1004.3239


 \bibitem{} Goldreich, P. 1966. ``Final spin states of planets and
            satellites." {\emph{The Astronomical Journal}}. Vol. {\bf{71}},
            pp. 1 - 7



 \bibitem{} Goldreich, P., and Peale, S. 1966. ``Spin-Orbit Coupling in the Solar System."
  {\it{The Astronomical Journal}}, Vol. {\bf{71}}, pp. 425 - 438


 \bibitem{}
 Hall, D.S. 1986.
 ``Pseudosynchronisation found in binaries with eccentric orbits."
 {\it{The Astrophysical Journal -- Letters to the Editor}}, Vol. {\bf{309}}, pp. L83 - L85

 \bibitem{}
 Hut, P. 1981.
 ``Tidal evolution in close binary systems." {\it{Astronomy \& Astrophysics}}, Vol. {\bf{99}}, pp. 126 - 140


 \bibitem{}
 K{\"{a}}pyl{\"{a}}, P. J., \& Brandenburg, A. 2008.
 ``Lambda effect from forced turbulence simulations."
 {\it{The Astrophysical Journal,}} Vol. {\bf{488}}, pp. 9 - 23

 \bibitem{} Karato, S.-i., and Spetzler, H. A. 1990. ``Defect Microdynamics
            in Minerals and Solid-State Mechanisms of Seismic Wave
            Attenuation and Velocity Dispersion in the Mantle."
            {\it{Reviews of Geophysics}}, Vol. {\bf{28}}, pp. 399 - 423


 \bibitem{} Kaula, W. M. 1964.  ``Tidal Dissipation by Solid Friction and
            the Resulting Orbital Evolution." {\it{Reviews of Geophysics}},
            Vol. {\bf{2}}, pp. 661 - 684

 \bibitem{}
 Kichatinov, L. L. 2005.
 ``The differential rotation of stars."
 {\it{Physics -- Uspekhi,}} Vol. {\bf{88}}, pp. 449 - 467


 \bibitem{}
 Makarov, V. V. 2012.
 ``Conditions of passage and entrampent of terrestrial planets in spin-orbit resonances."
 {\it{The Astrophysical Journal}}, Vol. {\bf{752}}, ~id. 73.\\
 doi:10.1088/0004-637X/752/1/73 ~~~~http://arxiv.org/abs/1110.2658

 \bibitem{} Makarov, V. V.; Berghea, C.; and Efroimsky, M. 2012.
 ``Dynamical evolution and spin-orbit resonances of potentially habitable exoplanets. The case of GJ 581d."
 {\it{The Astrophysical Journal}}, Vol. {\bf{761}}, ~id. 83.\\
 http://arxiv.org/abs/1208.0814\\
 ERRATUM: ApJ, Vol. {\bf{763}}, ~id. $\,$68


 \bibitem{} MacDonald, G. J. F. 1964. ``Tidal Friction." {\it{Reviews of
            Geophysics.}} Vol. {\bf{2}}, pp. 467 - 541

 \bibitem{} Miguel, M.-C.; Vespignani, A.; Zaiser, M.; and Zapperi, S. 2002.
          ``Dislocation Jamming and Andrade Creep." {\emph{Physical Review Letters}},
           Vol. {\bf{89}}, pp. 165501-1 - 165501-4.

 \bibitem{}
 Murray, C.D., and Dermott, S.F. 1999.
 {\it Solar System Dynamics}. Cambridge University Press, Cambridge UK


\bibitem{} Nakamura, Y.; Latham, G.; Lammlein, D.; Ewing, M.; Duennebier, F.; and Dorman, J. 1974.
            ``Deep lunar interior inferred from recent seismic data."
            {\it{Geophysical Research Letters}}, Vol. {\bf{1}}, pp. 137 - 140.

 \bibitem{} Ogilvie, G. I. 2008. ``James Clerk Maxwell and the dynamics of astrophysical discs." {\it{Philosophical Transactions of
 the Royal Society A}}, Vol. {\bf{366}}, pp. 1707 - 1715\\
 http://rsta.royalsocietypublishing.org/content/366/1871/1707.full.html


 \bibitem{}
 {Rudiger} R\"{u}diger, G. 1989.
 {\it Differential Rotation and Stellar Convection}. Gordon and Breach, NY

 \bibitem{}
 Segatz, M.; Spohn, T.; Ross, M.N.; and
 Shubert, G. 1988. ``Tidal dissipation, surface heat flow, and figure of viscoelastic models of Io."
 {\it{Icarus}}, Vol. {\bf{75}}, pp. 187 - 206

 \bibitem
 {sels} Selsis, F.; Wordsworth, R.D.; \& Forget, F. 2011. ``Thermal phase curves of nontransiting terrestrial exoplanets. I.
 Characterising atmospheres." {\it{Astronomy \& Astrophysics}}, Vol. {\bf{532}}, article id. A1

 \bibitem{}
 Strassmeier, K.G.; Carroll, T.A.; Weber, M.; Granzer, T.; Bartus, J.;
 Oláh, K.; and Rice, J. B. 2011. ``Binary-induced magnetic activity?. Time-series echelle spectroscopy and photometry of HD 123351 = CZ
 CVn." {\it{Astronomy \& Astrophysics}}, Vol. {\bf{535}}, id. A98

 \bibitem{} Torres, G.; Andersen, J.; and Giménez, A. 2010. ``Accurate masses and radii of normal stars: modern results and applications."
 {\it{The Astronomy and Astrophysics Review,}} Vol. {\bf{18}}, pp. 67 - 126

\bibitem{} Touma, J., and Wisdom, J. 1994. ``Evolution of the Earth-Moon
            system." {\emph{The Astronomical Journal.}} Vol. {\bf{108}},
            pp. 1943 - 1961.


 \bibitem{} Weber, R. C.; Lin, Pei-Ying; Garnero, E.; Williams, Q.; and Lognonn\'{e}, P. 2011. ``Seismic Detection of the Lunar Core."
            {\it{Science}}, Vol. {\bf{331}}, Issue 6015, pp. 309 - 312.



  \bibitem{} Williams, J. G., Boggs, D. H., and Ratcliff, J. T. 2008. ``Lunar Tides, Fluid Core and Core/Mantle Boundary."
            The 39th Lunar and Planetary Science Conference, (Lunar and Planetary Science XXXIX), held on 10-14 March 2008 in
            League City, TX. LPI Contribution No. 1391., p. 1484\\
            http://www.lpi.usra.edu/meetings/lpsc2008/pdf/1484.pdf

 \bibitem{} Williams, J. G., and Efroimsky, M. 2012. ``Bodily tides near the 1:1 spin-orbit resonance. Correction to Goldreich's
            dynamical model." {\it{Celestial Mechanics and Dynamical Astronomy}}, Vol. {\bf{114}}, pp. 387 - 414\\
            http://arxiv.org/abs/1210.2923

 \bibitem{} Williams, P. T. 2004. ``Turbulent magnetohydrodynamic elasticity: Boussinesq-like approximations for steady shear."
  {\it{New Astronomy}}, Vol. {\bf{10}}, p. 133-144\\
   http://arxiv.org/abs/astro-ph/0212556

 \bibitem{} Williams, P. T. 2005 ``Three routes to jet collimation by the Balbus-Hawley magnetorotational instability."
 {\it{Monthly Notices of the Royal Astronomical Society of London}}, Vol. {\bf{361}}, pp. 345 - 356\\
  http://arxiv.org/abs/astro-ph/0506184

 \bibitem{} Williams, P. T. 2006 ``Turbulent Elasticity of the Solar Convective Zone and the Taylor Number Puzzle." In:
 {\it{Solar MHD Theory and Observations: A High Spatial Resolution Perspective.}} Edited by H. Uitenbroek, J. Leibacher, and R. F. Stein.
 ASP Conference Series,  Vol. {\bf{354}}, pp. 85 - 91\\
 http://arxiv.org/abs/astro-ph/0602502 , ~~http://adsabs.harvard.edu/full/2006ASPC..354...85W

 \bibitem
 {word} Wordsworth, R.D.; Forget, F.; Selsis, F.; Madeleine, J.-B.; Millour, E.; and Eymet, V. 2010. ``Is Gliese 581d habitable?
 Some constraints from radiative-convective climate modeling." {\it{Astronomy \& Astrophysics}}, Vol. {\bf{522}}, article id. A22

 \end{thebibliography}
\end{document}